\documentclass[usenatbib]{mn2e}
\usepackage{natbibmnfix,graphicx,times,array}

\newcommand{\Msun}{\mathrm{M}_{\odot}}
\newcommand{\Map}{M_{\mathrm{ap}}}
\newcommand{\tout}{\theta_{\mathrm{out}}}
\newcommand{\Rout}{R_{\mathrm{out}}}
\newcommand{\dd}{\mathrm{d}}
\newcommand{\dNLlens}{\delta^{\mathrm{NL}}_l}
\newcommand{\rvir}{r_{\mathrm{vir}}}
\newcommand{\rhoBR}{\bar{\rho}}
\newcommand{\cvir}{c_{\mathrm{vir}}}
\newcommand{\Mpc}{\mathrm{Mpc}}
\newcommand{\Sigmacrit}{\Sigma_{\mathrm{crit}}}
\newcommand{\Sigmacritinf}{\Sigma_{\mathrm{crit},\infty}}
\newcommand{\finsideA}{\sqrt{\frac{1-x^2}{\cvir^2 - x^2}}}
\newcommand{\finsideB}{\sqrt{\frac{\cvir^2-x^2}{x^2 - 1}}}
\newcommand{\hyperR}{_2F_1\left[\frac{1}{2},\frac{\alpha}{2},\frac{3}{2},1-\frac{x_R^2}{x^2}\right]}
\newcommand{\hyperth}{_2F_1\left[\frac{1}{2},\frac{\alpha}{2},\frac{3}{2},1-\frac{\cvir^2}{x^2}\right]}

\onecolumn

\title[Lensing protoclusters]{The abundance of lensing protoclusters}

\author[D'Aloisio, Furlanetto \& Natarajan]{Anson
  D'Aloisio$^1$\thanks{Email: anson.daloisio@yale.edu},
  Steven R. Furlanetto$^2$ \& 
	Priyamvada Natarajan$^{1,3,4}$\\
$^1$Department of Physics, Yale University, PO Box 208120, New
    Haven, CT 06520-8120\\
$^2$Department of Physics and Astronomy, University of California, Los Angeles, CA 90095\\
$^3$Department of Astronomy, Yale University, PO Box 208101, New Haven, CT 06511\\
$^4$Radcliffe Institute for Advanced Study, Harvard University, 10 Garden Street, Cambridge, MA 02138}

\begin{document}

\maketitle

\begin{abstract}
Weak gravitational lensing provides a potentially powerful method for the detection of clusters.  In addition to cluster candidates, a large number of objects with possibly no optical or X-ray component have been detected in shear-selected samples.  Determining the nature of these so-called ``dark" lenses is an important step towards understanding the reliability of shear-selection techniques.  We develop an analytic model to investigate the claim of \citet{WK2002} that unvirialised protoclusters account for a significant number of dark lenses.  In our model, a protocluster consists of a small virialised region surrounded by in-falling matter.  We use a simple model for the density profile that assumes the Navarro-Frenk-White form inside of the virial radius and a power law $\rho \sim r^{-\alpha}$ outside.  We find that, in order for a protocluster to simultaneously escape X-ray detection and create a detectable weak lensing signal, it must have a small virial mass ($\sim 10^{13}~\Msun$) and large total mass ($\sim10^{15}~\Msun$), with a relatively flat density profile outside of the virial radius ($\alpha \sim 0 - 1$).  Such objects would be characterized by rising tangential shear profiles well beyond the virial radius.  We use a semi-analytic approach based on the excursion set formalism to estimate the abundance of lensing protoclusters with a low probability of X-ray detection.  We find that they are extremely rare, accounting for less than $0.4$ per cent of the total lenses in a survey with background galaxy density $n = 30 ~ \mathrm{arcmin^{-2}}$ and intrinsic ellipticity dispersion $\sigma_{\epsilon} = 0.3$.  Their  abundance decreases significantly if flat density profiles outside of the virial radius are not common.  We conclude that lensing protoclusters with undetectable X-Ray luminosities are too rare to account for a significant number of dark lenses.    
\end{abstract}


\section{INTRODUCTION}

The abundance of collapsed dark matter haloes in the Universe yields a wealth of information on the dynamics of structure formation.  The number density of galaxy clusters and its time-evolution can be used to probe the normalization of the linear power spectrum, $\sigma_8$, and the density parameters $\Omega_m$ and $\Omega_{\Lambda}$ \citep[e.g.,][]{Lilje1992,WEF1993,CenOstriker1994,ECF1996,Henry1997,BahcallFan1998,VianaLiddle1999,Holder2001,ReiprichBohringer2002,Dahle2006}.  In addition, since the linear growth of overdensities and comoving volume-element are sensitive to the dark energy equation-of-state parameter $w = P/\rho$, the cluster abundance can be used to constrain dark energy \citep[e.g.,][]{Haiman2001,MajumdarMohr2004,Mantz2008}.  In order to effectively use the cluster number counts as a cosmological tool, one must be able to create a complete catalog out to high redshifts.  To date, most studies aiming to complete this task have relied on X-ray based selection and constraints on cluster masses \citep[see however][]{Dahle2006}.  Mass estimates derived from their X-ray emission require the additional assumption of hydrostatic equilibrium for the gas, which is not robust.  

Galaxy clusters are the most recently assembled structures in the Universe.  One should therefore expect to find a large number of them in a dynamically unrelaxed state.  In recent years, much effort has been devoted to developing techniques that utilize weak-lensing in cluster surveys.  Gravitational lensing offers a method to measure masses that is independent of their dynamical state.  In addition to improving the accuracy of mass measurements, the weak distortion in the shapes of background galaxies may also provide a powerful method of cluster detection.  In this spirit, \citet{Schneider1996} introduces the aperture mass measure as a way to systematically search for mass concentrations using weak lensing.  In this approach, a weighted sum of image ellipticities is used as a proxy for the projected mass contained within an aperture. 

Since gravitational lensing probes mass concentrations in a way that is independent of their baryonic content, shear-selected samples can potentially provide a new and exciting view of large-scale structure.  In fact, there have been several shear-selected cluster candidates that appear to lack the characteristic galaxy over-density and X-ray luminosity. \citep[e.g.,][]{Erben2000,UmetsuFutamase2000,Miralles2002,Dahle2003,Schirmer2007,Maturi2007}.  The first detection of these so-called dark lenses was reported by \citet{Erben2000}.  Their initial analysis indicates significant tangential alignment roughly 7 arcminutes South of the cluster Abell 1942.  Assuming the presence of a mass concentration with roughly the same redshift as Abell 1942 ($z \sim 0.2$), they obtain a lower-bound mass estimate of $M \sim 10^{14} h^{-1}~\Msun$ inside of a sphere of radius $r = 0.5h^{-1}~\Mpc$.  However, a follow-up analysis by \citet{vonderLinden2006} using HST observations detects the tangential alignment with a lower significance, casting doubt on the hypothesis that the object is a true mass concentration.  The HST data contains a larger number of distant galaxies and should therefore increase the significance of the detection in the case of a dark mass concentration.  

Among the most recent detections, \citet{Schirmer2007} find 86 dark lenses out of 158 possible mass concentrations in a wide field with average galaxy density $n \sim 12~\mathrm{arcmin}^{-2}$.  These objects, which show no detectable optical component, were identified using the aperture mass measure and a variant of it.  Similarly, using the linear filter of \citet{Maturi2005} to minimize spurious signals created by large-scale structure (LSS), \citet{Maturi2007} detect 7 dark lenses with no optical or X-ray component out of 14 identified lenses.  The exact nature of these detections remains an open question.

There are several possible explanations for the appearance of dark lenses in shear-selected surveys.  One possibility is that these objects truly correspond to dark matter concentrations that are abnormally deficient in baryons.  A significant abundance of such objects is not expected and would require a rethinking of current structure formation scenarios \citep{vonderLinden2006,Maturi2007}.  A second possibility is that dark lenses are spurious signals in a lensing map resulting from the alignment of intrinsic galaxy ellipticities or LSS projected along the line-of-sight.  Both scenarios are likely to be significant problems for weak-lensing surveys and have been investigated by several authors \citep{ReblinskyBartelmann1999,White2002,Hamana2004,Hennawi2005,Pace2007,Fan2007}. 

Another interesting possibility was proposed by \citet{WK2002} (WK2002).  They suggest that dark lenses may be cluster progenitors that are not fully virialised.  They argue that these objects should have a low galaxy over-density and X-ray luminosity compared to fully virialised clusters of the same mass.  If some of these objects are sufficiently massive and over-dense to create a detectable weak-lensing signal, then they might have the same observable signatures as dark lenses.  For clarity, we will refer to these objects as lensing protoclusters (LP).  Using an analytic approach based on the Press-Schechter formalism, WK2002 estimate that $10-20$ per cent of weak lenses should be dark LPs.                         

Of course, it is entirely possible that the dark lens phenomenon is due to a combination of the above scenarios.  Determining the extent to which each contributes to dark lens abundances, if at all, is an important step towards understanding the reliability of shear- selection techniques.  In this paper, we explore the scenario proposed by WK2002 in further detail.  We create a simple analytic model to investigate the likelihood that LPs can have the observational properties of dark lenses. 

The remainder of this paper is organized in the following manner.  In section \ref{APMASSMEASURE}, we briefly review the aperture mass technique.  In section \ref{AMLPs}, we present our analytic model of LPs.  In section \ref{LPsDL}, we calculate the properties that LPs must have to be detected as dark lenses.  We calculate the dark LP mass function in section \ref{ABUNDANCES} and explore what the abundance of dark LPs implies for weak-lensing surveys.  Finally, we discuss our results and other potential dark lenses in section \ref{DISCUSSION}.

In what follows, we assume a $\Lambda$CDM cosmology with parameters $\Omega_m = 0.26$, $\Omega_{\Lambda} = 0.74$, $\Omega_b = 0.044$, $H = 100h~\mathrm{km s^{-1}}~\Mpc^{-1}$ (with $h = 0.72$), $n = 0.96$, and $\sigma_8 = 0.8$, consistent with five-year WMAP results \citep{Dunkley2008}.  Unless otherwise noted, all distances and volumes are reported in physical units.
  
  \section{The aperture mass measure} \label{APMASSMEASURE}

We use the aperture mass signal-to-noise ratio to define a detectable weak lensing signal in a shear-selected cluster survey.  In this section, we briefly review the aperture mass technique \citep{Schneider1996}.

In order to realistically model a survey's sensitivity to mass concentrations, a functional form for the redshift distribution of background galaxies in accordance with observations needs to be assumed.  Following the analysis of WK2002, we assume a source redshift distribution \citep{Brainerd1996},

\begin{equation}
p_z(z_s) = \frac{\beta z_s^2}{\Gamma(3/\beta)z_0^3}~\exp\left[ -(z_s/z_0)^\beta \right],
\end{equation}
with a mean redshift of 1.2 ($z_0 = 0.8$ and $\beta = 1.5$).  In what follows we will find it convenient to isolate the source redshift dependence of all physical quantities by defining a redshift weight function \citep{Seitz1997,WK2002}, 

\begin{equation}
Z(z_s;z_l) \equiv \frac{\lim_{z_s \rightarrow \infty}{\Sigmacrit(z_l;z_s)}}{\Sigmacrit(z_l;z_s)}H(z_s - z_l) \equiv\frac{\Sigmacritinf(z_l)}{\Sigmacrit(z_l;z_s)},
\label{Zeq}
\end{equation}
where $z_s$ and $z_l$ are the source and lens redshifts respectively, $\Sigmacrit$ is the critical surface mass density, and $H$ is the Heaviside step function.  Note that the $z_s$-dependence of the convergence, $\kappa = \Sigma/\Sigmacrit$, where $\Sigma$ is the surface mass density, can be factored out by writing $\kappa(\vec{\theta},z_s,z_l) = \kappa_{\infty}(\vec{\theta},z_l)Z(z_s;z_l)$, where $\kappa_{\infty}(\vec{\theta},z_l) \equiv \Sigma(\vec{\theta})/\Sigmacritinf(z_l)$ is interpreted to be the convergence for the case of reference sources at $z_s = \infty$.  Similarly, the shear can be written as $\gamma(\vec{\theta},z_s,z_l) = \gamma_{\infty}(\vec{\theta},z_l)Z(z_s;z_l)$.  We will exclusively use the $z_s$-independent versions of the convergence and shear from here on.   
 
The aperture mass provides a way to detect mass concentrations through the distortion of lensed background images.  It is defined as a weighted integral over the convergence,

\begin{equation}
M_{\mathrm{ap}}(\vec{\theta_0}) = \int{d^2\theta~\kappa(\vec{\theta}~)~U(|\vec{\theta} - \vec{\theta_0}}|),
\label{Map1}
\end{equation}
where $U(|\vec{\theta} - \vec{\theta_0}|)$ is a compensated weight function centered at $\vec{\theta_0}$, satisfying $\int{\dd \theta~\theta~U(\theta)} = 0$.  As desired, $\Map$ can be expressed in terms of the tangential component of the shear relative to the direction $\vec{\theta}_0$, $\gamma_t \equiv -\Re[\gamma(\vec{\theta}+\vec{\theta_0})\exp(-2i\phi)] $.  Using the so-called Kaiser-Squires inversion \citep{KaiserSquires1993}, one can obtain

\begin{equation}
M_{\mathrm{ap}}(\vec{\theta_0}) = \int{d^2 \theta
~\gamma_t(\vec{\theta};\vec{\theta_0})~Q(|\vec{\theta}-\vec{\theta_0}|)},
\label{Map2}
\end{equation}
where $Q(\theta) = \frac{2}{\theta^2}\int_0^{\theta}{dx~x U(x)} - U(\theta)$.

Equations (\ref{Map1}) and (\ref{Map2}) illustrate how the tangential shear field can be used as a measure of the integrated mass within an aperture.  However, in practice, a discrete analog to equation (\ref{Map2}) in terms of a sum over individual image ellipticities is more useful:
 
\begin{equation}
M_{\mathrm{ap}}(\vec{\theta_0}) = \frac{1}{n} \sum_i{Q(|\vec{\theta_i} - \vec{\theta_0}|)~\epsilon_{ti}(\vec{\theta_i};\vec{\theta_0})}.
\label{discMapEQ}
\end{equation}
Here, $n$ is the number density of galaxy images and $\epsilon_{ti}(\vec{\theta_i};\vec{\theta}_0)$ is the tangential component of the ellipticity relative to $\vec{\theta}_0$.

Setting $\vec{\theta}_0 = 0$ for simplicity, the expectation value of $\Map$ is given by

\begin{equation}
\left< \Map \right> = \frac{\left< Z \right>}{n}\sum_i{Q(|\vec{\theta}_i|)\gamma_{t,\infty}(\vec{\theta}_i)},
\label{avgMapEQ}
\end{equation}
where $\left< Z \right> = \int_0^{\infty}{dz_s~p_z(z_s)Z(z_s;z_l)}$.  Note that in obtaining equation (\ref{avgMapEQ}), we have used $\left< \epsilon_t \right> \approx \left< Z \right>\gamma_{t,\infty}$ in the weak lensing regime.  In the case with no lensing, $\left< \Map \right> = 0$, the dispersion of $\Map$ is obtained by squaring equation (\ref{discMapEQ}) and taking the expectation value.  Assuming that tangential ellipticities of individual galaxies are uncorrelated, $\left< \epsilon_{ti}~\epsilon_{tj} \right> = 0$ for $i \neq j$, the dispersion is

\begin{equation}
\sigma^2_{\mathrm{ap}} = \frac{\sigma^2_{\epsilon}}{2 n^2} \sum_{i}{Q^2(|\vec{\theta}_i|)},
\label{sigEQ}
\end{equation}
where $\sigma_{\epsilon}$ is the dispersion of intrinsic galaxy ellipticities.

Averaging equations (\ref{avgMapEQ}) and (\ref{sigEQ}) over the probability distribution of galaxy positions and taking their ratio yields the ensemble averaged signal-to-noise ratio,

\begin{equation}
\frac{S}{N} = \frac{2 \left< Z \right> \sqrt{\pi n}}{\sigma_{\epsilon}} 
	\frac{\int_0^{\tout}{\dd \theta~\theta~\left< \gamma_{t,\infty} \right>(\theta)~Q(\theta)}}{\sqrt{\int_0^{\tout}{\dd \theta~\theta~ Q^2(\theta)}}}
\label{StoNEQ}
\end{equation}
where $\tout$ is the angular radius of the aperture and $\left< \gamma_{t,\infty} \right>(\theta)$ is the average tangential shear on a circle of angular radius $\theta$.  Using the Cauchy-Schwarz inequality the maximum signal-to-noise ratio is obtained by selecting $Q(\theta) \propto \left< \gamma_t \right>(\theta)$.  This fact is intuitively clear; the signal is maximized by choosing the shear profile as its own weight function.

In this paper, we use the weight function developed by \citet{Schirmer2004} \citep[see also][]{Schirmer2007},

\begin{equation}
Q(\theta) = E(X)~\frac{\tanh(X/x_c)}{X/x_c}
\label{ApertureEQA}
\end{equation}
where

\begin{equation}
E(X) = \frac{1}{1+\mathrm{e}^{6 - 150X}+\mathrm{e}^{-47+50X}}.
\label{ApertureEQB}
\end{equation}
Here, $X \equiv \theta/\tout$ and $x_c$ is a dimensionless width parameter.  Smaller $x_c$ results in more weight towards small radii.  The above filter was developed as a computationally inexpensive alternative to using the NFW shear profile.  It is designed to optimally detect NFW haloes in a wide field survey.  Since our goal is to determine whether LPs are a significant contaminant in cluster surveys, the above filter will provide a more realistic estimate of their signal-to-noise.

In what follows we use $S/N = 4$ as our shear-selection threshold \citep[see][for example]{Schirmer2007}.  We adopt the fiducial values of $n = 30~ \mathrm{arcmin}^{-2}$ and $\sigma_{\epsilon} = 0.3$.  In practice, a wide variety of aperture radii are used in order to detect clusters of different scales and at various redshifts.  Since we aim to detect extended objects that are still in the process of collapsing, we adopt an aperture size of $R_{\mathrm{out}} = \tout D_A(z_l) = 5~\Mpc$, where $D_A(z_l)$ is the angular diameter distance to the lens.  This form is ideal for detecting objects with a scale of $\sim 5~\Mpc$ at any redshift.  Note that $\tout$ becomes redshift dependent and corresponds to using $\tout = 26$, $14$, $12$, and $11$ arcminutes for LPs at redshifts of $z = 0.2$, $0.5$, $0.7$, and $0.9$ respectively.  

\section{Analytic model of lensing protoclusters} \label{AMLPs}

\subsection{The LP mass profile}

We qualitatively define LPs to be progenitors of cluster-scale haloes that are not fully virialised.  Since haloes gain mass by accretion along their outskirts, it is natural to suspect that a protocluster consists of a small region in virial equilibrium surrounded by an infall envelope.  We can therefore separate the protocluster into two distinct parts: the central virialised region (CVR) and the infall region (IF).  In what follows, we define the virial radius $\rvir$ such that the average over-density inside of $\rvir$ is $\ge 200$ times the critical density of the universe at that epoch - a convention frequently used in N-body simulations.  As \citet{Cuesta2008} point out, this definition does not provide a robust approximation to the true virial radius.  We nonetheless adopt it for its simplicity and convenience in comparison to other works that have used this convention.  Using different definitions of $\rvir$ does not change the main conclusions of this paper.  

We model the mass density inside of the CVR with the ubiquitous Navarro-Frenk-White (NFW) profile,

\begin{equation}
\rho(r) = \frac{\rho_s}{(r/r_s)(1+r/r_s)^2},
\label{NFWprofile}
\end{equation}
where $\rho_s$ and $r_s$ are free parameters  \citep{NFW1995,NFW1996,NFW1997}.  \citet{Tavio2008} have shown that (\ref{NFWprofile}) provides a good fit to density profiles \emph{within the virial radius}. It is often convenient to characterize the above profile with the concentration parameter, $\cvir = \rvir/r_s$.  By integrating equation (\ref{NFWprofile}) out to $\rvir$ and using $m_v = 200\rho_c(z)~4\pi \rvir^3/3$, where $m_v$ is defined to be the virial mass and $\rho_c$ is the critical density of the universe, the concentration parameter can be related to $\rho_s$ through

\begin{equation}
\rho_s = \frac{200\rho_c(z)}{3}\frac{\cvir^3}{\ln(1+\cvir) - \cvir/(1+\cvir)}.
\label{rhoseq}
\end{equation}
The concentration depends on the virial mass and redshift of the halo under consideration.  In this paper, we use the form \citep{Seljack2000,TakadaJain2002}, 
\begin{equation}
\cvir(m_v,z) = \frac{10}{1+z}~\left( \frac{m_v}{M_*(z=0)} \right)^{-0.2},
\label{concentrationeq}
\end{equation}
where $M_*(z=0)$ is the present day non-linear mass scale ($\delta_c(z = 0)/\sigma(M_*) = 1$).  As \citet{TakadaJain2002} point out, the halo model using the above form is known to be in better agreement with the non-linear matter power spectrum compared to other choices \citep{Seljack2000,Cooray2000}.

We turn our attention to a quantitative description of the infall envelope surrounding the central regions of a LP.  The outskirts are likely anisotropic due to the fact that accretion occurs along filamentary structures as observed in numerical simulations.  Moreover, since the infall region consists of smaller haloes, we expect sub-structure to play an important role.  However, the incorporation of both of these characteristics is beyond the scope of our analytic model. Instead, we content ourselves with developing a spherically symmetric profile describing regions beyond $\rvir$.  

In what follows, we assume that the density profiles of LPs do not fall off as steeply as the NFW profile for $r > \rvir$; we model the infall regions with a power law $\rho \sim r^{-\alpha}$, where $\alpha < 3$.  This choice is motivated by several results. Using N-body simulations, \citet{ENF1998} find a significant deviation from the NFW form at large radii, especially at higher redshifts (see Figure 10 of their paper).  Secondly, the excursion set formalism can be used to show that the average density within infall regions falls off more slowly than $r^{-3}$  \citep{Barkana2004}.  Most recently, it has been pointed out that the NFW form provides a poor fit outside of the virial radius \citep{Prada2006,Cuesta2008,Tavio2008}.  Using N-body simulations, \citet{Tavio2008} develop a density profile that better describes these regions on average.  Their form closely approximates the NFW profile for small radii, but the instantaneous logarithmic slopes typically transition from $\sim -3$ at $\rvir$ to $\sim -0.2$ at $10\rvir$. They also find large variations in the density profiles of individual haloes beyond $2\rvir$.  Hence, rather than determine $\alpha$ through dynamical arguments, we will explore what types of infall profiles are required to produce a detectable weak lensing signal in section \ref{LPsDL}. 
   
Following the above discussion, we model LPs with the density profile

\begin{equation}
\rho(r) =  \left\{ \begin{array}{ll} \frac{\rho_s}{r/r_s (1 + r/r_s)^2} & \mbox{$r \le \rvir$} \\ 
	 \frac{\rho_0}{(r/\rvir)^{\alpha}} & \mbox{$\rvir < r \le R$} \\ 
	 0 & \mbox{$R < r$} \end{array} \right.
\label{LPprofileEQ}
\end{equation}
where $\rvir$ is the virial radius of the CVR and $R$ is a truncation radius, introduced to keep the mass of the profile finite.  Throughout the rest of this paper, we will refer to equation (\ref{LPprofileEQ}) as the LP profile.  

Note that (\ref{LPprofileEQ}) is uniquely determined by four parameters: $M$ (the total mass inside of the truncation radius $R$), $m_v$ (the virial mass), $z_l$ (the redshift of the LP), and $\alpha$ (the logarithmic slope of the density profile outside $\rvir$).  Given $m_v$ and $z_l$, one can immediately obtain $\rvir$ using $m_v = 200\rho_c(z)~4\pi \rvir^3/3$.  Equations (\ref{concentrationeq}) and (\ref{rhoseq}) may then be used to obtain $\cvir$ and $\rho_s$ respectively.  Finally, $\rho_0$ is obtained by imposing continuity at $r = \rvir$.

\subsection{The surface mass density and tangential shear}

\begin{figure}
\begin{center}
\resizebox{8.0cm}{!}{\includegraphics{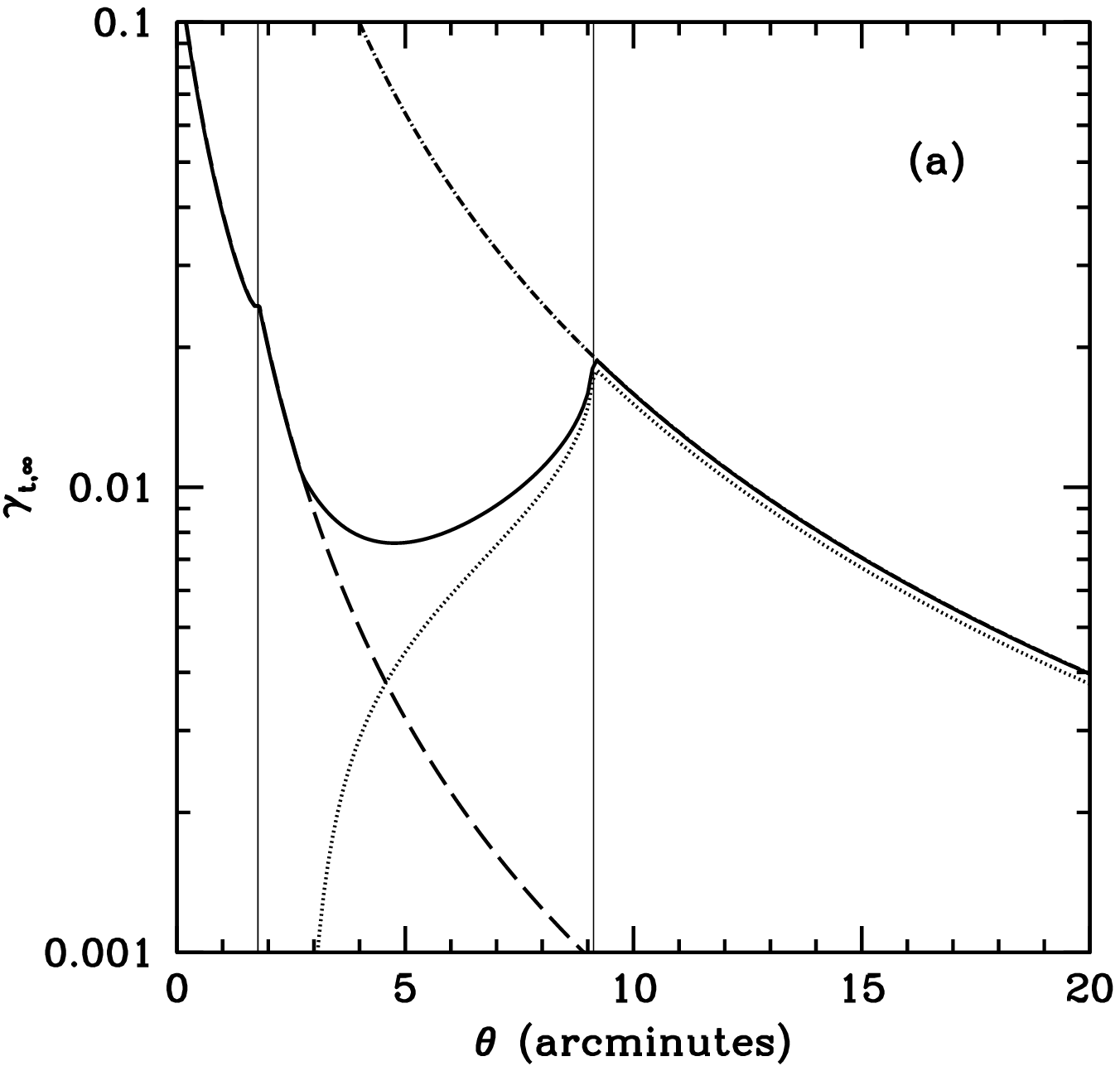}} \hspace{0.13cm}
\resizebox{8.0cm}{!}{\includegraphics{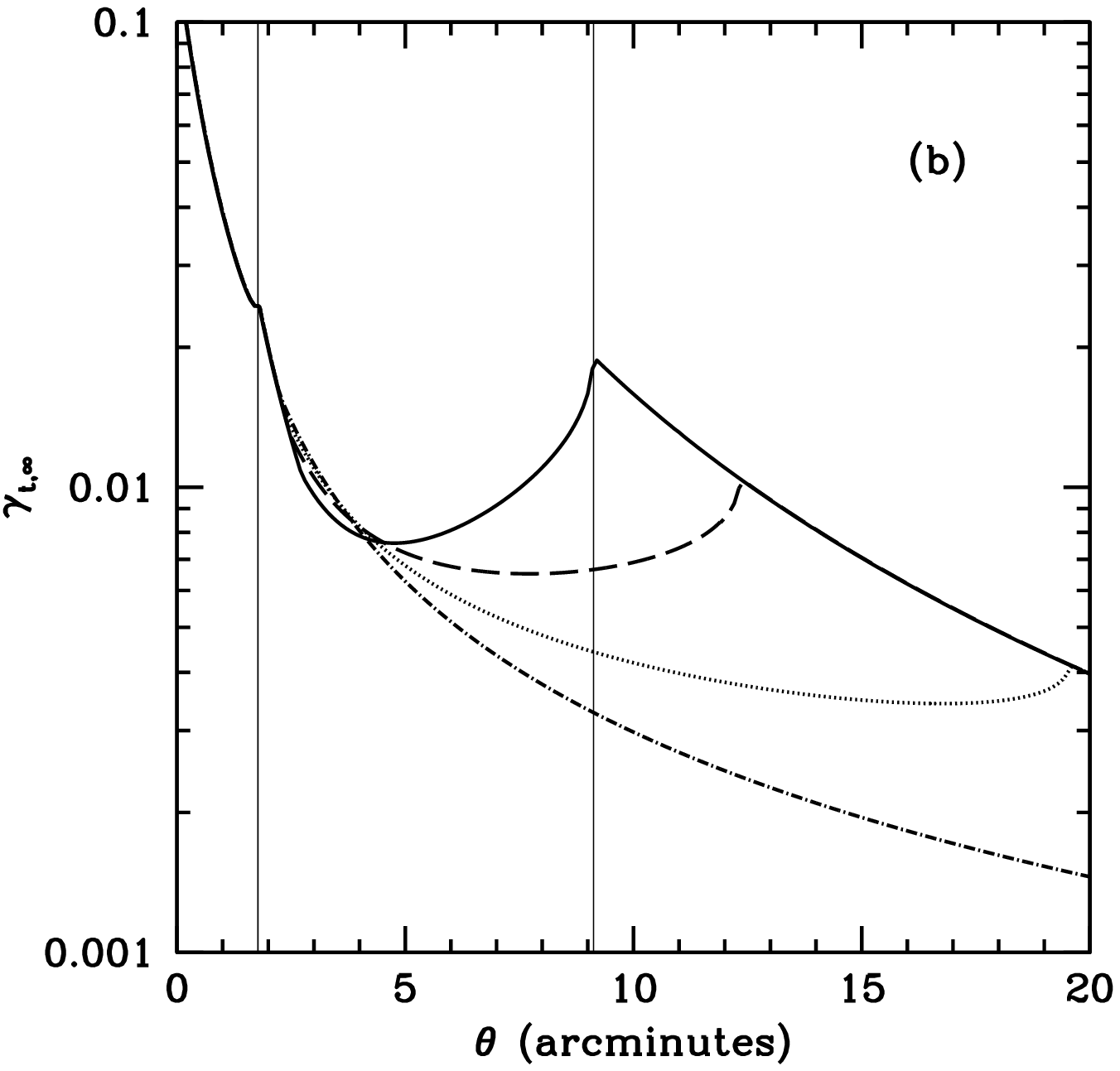}}
\end{center}
\caption{Panel $(a)$:  the solid line shows the tangential shear of sources at $z_s = \infty$ induced by the LP profile.  We assume parameters of $M = 10^{15}~\Msun$, $m_v = 5\times 10^{13}~\Msun$, $\alpha = 0.5$ and $z_l = 0.5$.  The dashed and dotted lines show the contributions from the CVR and infall region respectively.  The dot-dashed line corresponds to a point mass of mass $M$.  The tangential shear corresponding to the LP is the sum of contributions from the CVR and infall regions.  Panel $(b)$:  the tangential shear due to the LP profile for values of $\alpha = 0.5$ (solid), $1$ (dashed), $1.5$ (dotted), $2$ (dot-dashed).  All other parameters are the same as in $(a)$.  Protocluster density profiles that fall off more slowly with radius result in steeper up-turns in the tangential shear.}
 \label{TanShearPL}
 \end{figure}

Let us define a lens-centered coordinate system, $\left\{r_1,r_2,r_3 \right\}$, such that $r_3$ is along the line-of-sight to the lens center.  Our first task is to calculate the surface mass density, $\Sigma(r_1,r_2) = \int_{-\infty}^{\infty}{dr_3~\rho(\vec{r}~)}$, of the LP profile in the 2 regions: the CVR and the outer infall region.  Defining 

\begin{equation}
\setlength{\extrarowheight}{0.5cm}
f(x) = \left\{ \begin{array}{ll} \frac{\sqrt{\cvir^2-x^2}}{(x^2-1)(1+\cvir)} - \frac{\tanh^{-1}\left[\finsideA\right]}{(1-x^2)^{3/2}} + \frac{\tanh^{-1}\left[\cvir \finsideA\right]}{(1-x^2)^{3/2}} & \mbox{$x<1$} \\ 
	 \frac{ 2 + \cvir (\cvir^2 - 3)} {3(\cvir^2-1)^{3/2}} & \mbox{$x=1$} \\ 
 \frac{\sqrt{\cvir^2-x^2}}{(x^2-1)(1+\cvir)} - \frac{\tan^{-1}\left[\finsideB\right]}	{(x^2-1)^{3/2}} + \frac{\tan^{-1}\left[\frac{1}{\cvir}\finsideB\right]}{(x^2-1)^{3/2}} & \mbox{$x>1$}, \end{array} \right.	
\label{fofx}
\end{equation}
we obtain

\begin{equation}
\Sigma(x)_{\mathrm{LP}} = \Sigma(x)_{\mathrm{CVR}}+\Sigma(x)_{\mathrm{IF}},
\label{SMDuclEQ}
\end{equation}
where

\begin{equation}
\Sigma(x)_{\mathrm{CVR}} = \left\{ \begin{array}{ll} 2 \rho_s r_s f(x) & \mbox{$x < \cvir$} \\ \cr 0 & \mbox{$ x \ge \cvir$} \end{array} \right.
\label{sigvirEQ}
\end{equation} 
is the surface mass density of a NFW profile truncated at $\rvir$ and 

\begin{equation}
\Sigma(x)_{\mathrm{IF}} = \left\{ \begin{array}{ll} 2\rho_0 r_s \cvir^{\alpha}~x^{-\alpha}  \left\{ \sqrt{x_R^2-x^2} ~\hyperR - \sqrt{\cvir^2-x^2}~\hyperth \right\} & \mbox{$x < \cvir$} \\ \cr
	 2 \rho_0 r_s \cvir^{\alpha}~ x^{-\alpha} \sqrt{x_R^2 - x^2}~\hyperR & \mbox{$ \cvir \le x < x_R$} \\ \cr 
	 0 & \mbox{$x \ge x_R$}. \end{array} \right.
\label{ZeropowSurfMass}
\end{equation}
is the contribution from the infall profile, 
\begin{equation}
\rho(r) = \left\{ \begin{array}{ll} 0 & \mbox{$r < \rvir$} \\ 
	 \frac{\rho_0}{(r/\rvir)^{\alpha}} & \mbox{$\rvir \le r < R$} \\ 
	 0 & \mbox{$R \le r$}. \end{array} \right.
\label{perimeterprof}
\end{equation}  
Here, $_2F_1$ are hypergeometric functions, $x \equiv (1/r_s)\sqrt{r_1^2 + r_2^2}$ is the projection of the coordinate vector in the plane perpendicular to the line of sight in units of $r_s$, and $x_R = R/r_s$ is the dimensionless truncation radius.

The fact that equation (\ref{SMDuclEQ}) is a sum of contributions from the CVR and infall regions will prove to be extremely helpful when quantifying the lensing contribution from unvirialised matter in section \ref{APMSTON}.  The $z_s$-independent convergence is obtained from (\ref{SMDuclEQ}) using $\kappa_{\infty} = \Sigma/\Sigmacritinf$.  The corresponding average shear profile, $\left< \gamma_{t,\infty} \right>(\theta)$, can be calculated using $\left< \gamma_{t,\infty} \right> = \bar{\kappa}_{\infty}(\theta) - \left< \kappa_{\infty} \right>(\theta)$, where $\bar{\kappa}_{\infty}(\theta)$ and $\left< \kappa_{\infty} \right>(\theta)$ are the average value of the convergence inside and on a circle of angular radius $\theta$ respectively.  Owing to spherical symmetry, $\left< \kappa_{\infty} \right>(\theta) = \kappa_{\infty}(\theta)$ and $\left< \gamma_{t,\infty} \right>(\theta) = \gamma_{t,\infty}(\theta)$ for the LP profile. 
  
We now investigate the tangential shear of sources at $z_s = \infty$ induced by the LP profile.  The solid curve in Figure \ref{TanShearPL} $(a)$ shows $\gamma_{t,\infty}$ with lens parameters $M = 10^{15}~\Msun$, $m_v = 5 \times 10^{13}~\Msun$, $z_l = 0.5$, and $\alpha = 0.5$.  The left and right vertical lines represent the virial and truncation radii respectively.  Here, $\rvir = 0.64~\Mpc$ and $R = 3.3~\Mpc$, corresponding to $\theta = 1.8$ and $9.1$ arcminutes respectively.  Note that there are two kinks in the tangential shear.  These kinks occur because (\ref{LPprofileEQ}) is an idealized density profile, with sharp boundaries at $\rvir$ and $R$.  The dashed and dotted curves in panel $(a)$ correspond to contributions to the shear from the CVR and infall region respectively. The LP tangential shear profile is the sum of these contributions.  For reference, we also show the tangential shear induced by a point mass with $M = 10^{15}~\Msun$ (dot-dashed).  The plot shows that the LP $\gamma_{t,\infty}$ is well approximated by the CVR $\gamma_{t,\infty}$ inside of $\rvir$.  The LP $\gamma_{t,\infty}$ rises outside of $\rvir$, where the infall region contributes more to the shear.  Outside of the truncation radius, the tangential shear is equivalent to the case of a point mass with the same mass $M$.    

In Figure \ref{TanShearPL} $(b)$, we vary the power-law index $\alpha$ for fixed values of $M = 10^{15}~\Msun$, $m_v = 5 \times 10^{13}~\Msun$ and $z_l = 0.5$.  The solid, dashed, dotted, and dot-dashed curves correspond to $\alpha = 0.5$, $1$, $1.5$, and $2.0$ respectively.  The rise in $\gamma_{t,\infty}$ is most pronounced for flatter power laws.

\subsection{The aperture mass signal-to-noise ratio} \label{APMSTON}

\begin{figure}
\begin{center}
\resizebox{8.0cm}{!}{\includegraphics{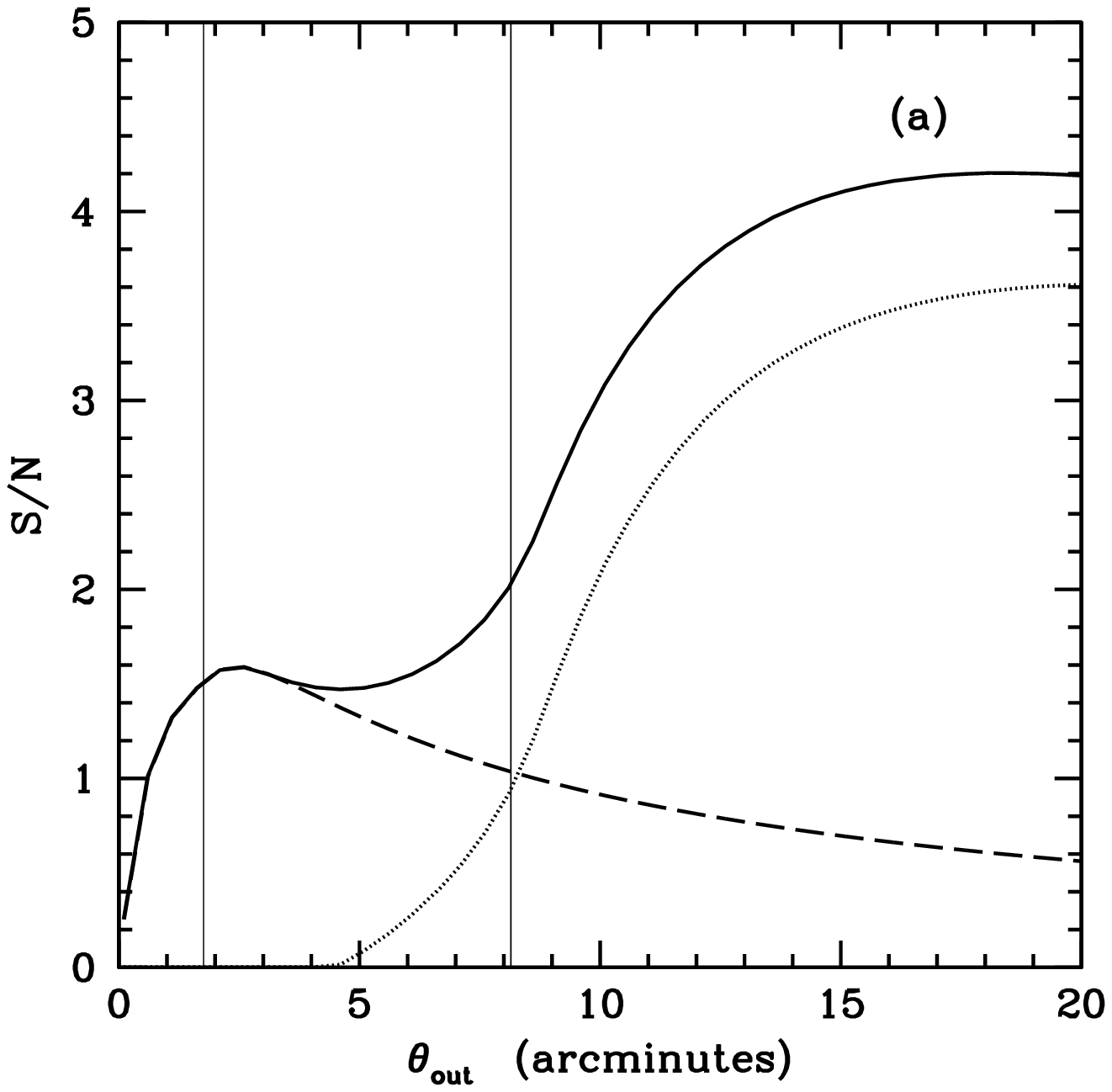}} \hspace{0.13cm}
\resizebox{8.0cm}{!}{\includegraphics{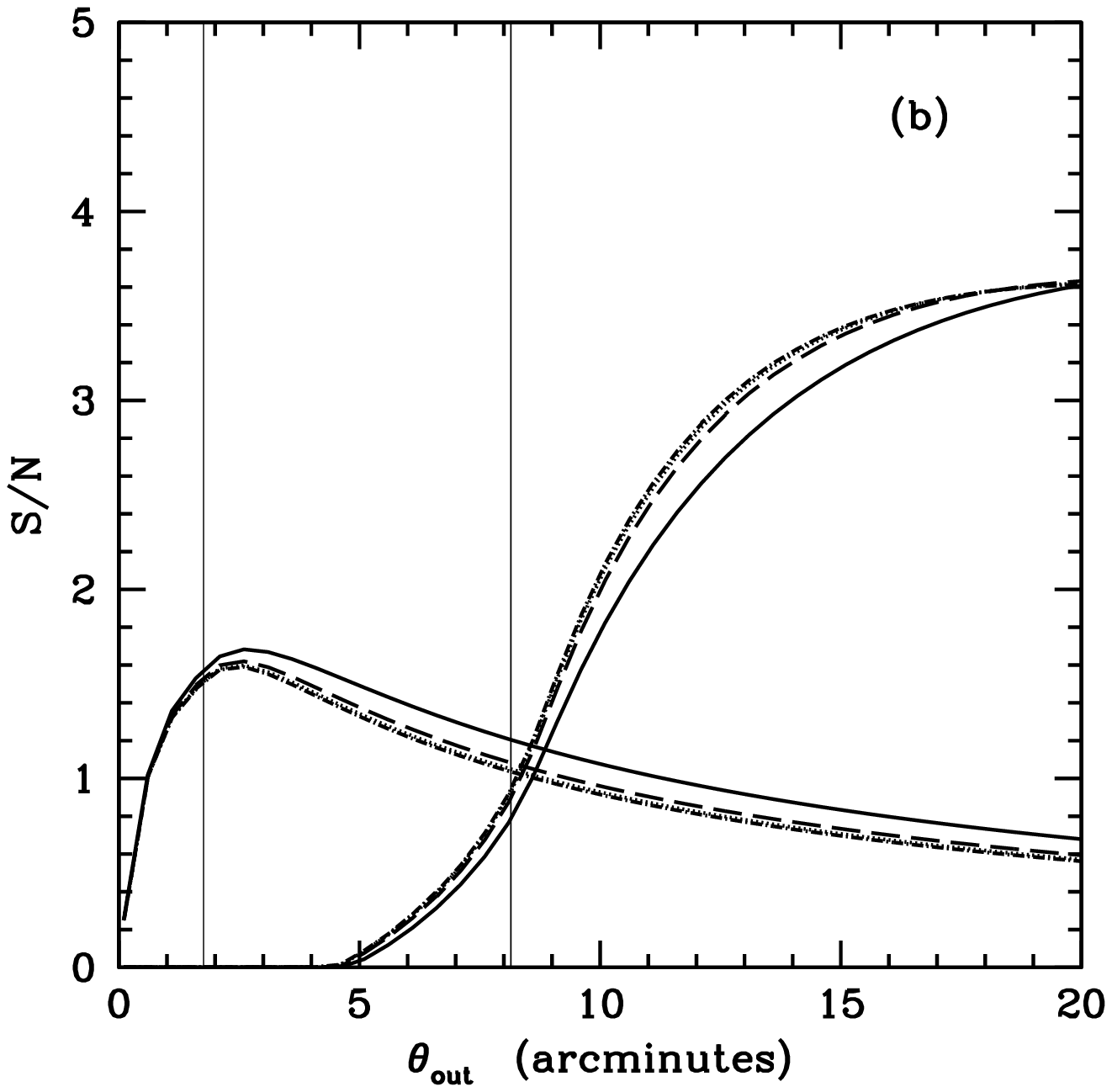}}
\end{center}
\caption{Panel $(a)$:  the aperture mass signal-to-noise ratio as a function of aperture radius $\tout$.  The solid, dashed, and dotted lines correspond to the LP profile and contributions from the CVR and infall regions respectively.  We assume an aperture width parameter of $x_c = 2$, $M = 10^{15}~\Msun$, $m_v = 5\times10^{13}~\Msun$, and $\alpha = 0.29$ (chosen so the $S/N = 4$ for $\tout = 14$ arcminutes).  Panel $(b)$: same as $(a)$ but with values of $x_c = 0.5$ (solid), $1$ (dashed), $1.5$ (dotted), and $2$ (dot-dashed).}
 \label{StoNtoutPL}
 \end{figure}

In this section we investigate the signal-to-noise properties of the LP.  The solid line in Figure \ref{StoNtoutPL} $(a)$ shows the LP signal-to-noise ratio as a function of aperture radius for $z_l = 0.5$, $M = 10^{15}~\Msun$, $m_v = 5 \times 10^{13}~\Msun$, and $\alpha = 0.26$ (chosen so that $S/N = 4$ for $\Rout = 5~\Mpc$, corresponding to $\tout = 14$ arcminutes).  The left and right vertical lines correspond to the virial and truncation angular radii respectively.  Since equation (\ref{StoNEQ}) is linear in $\left< \gamma_{t,\infty} \right>$, the LP signal-to-noise ratio is the sum of contributions from the CVR and infall profiles.  The dashed and dotted lines show the corresponding CVR and infall signal-to-noise ratios.  Figure \ref{StoNtoutPL} illustrates that, for smaller aperture radii, the LP signal-to-noise is dominated by the CVR.  On the other hand, the infall envelope makes a significant contribution in larger apertures.

The left and right curves in Figure \ref{StoNtoutPL} $(b)$ show $(S/N)_{\mathrm{CVR}}$ and $(S/N)_{\mathrm{IF}}$ using the same profile parameters as above.  The solid, dashed, dotted and dot-dashed curves correspond to aperture width parameters of $x_c = 0.5$, $1.0$, $1.5$, and $2.0$ respectively.  Since lower values of $x_c$ result in more weight towards smaller radii, the CVR contribution increases as $x_c$ decreases.  Conversely, the infall contribution is suppressed as $x_c$ decreases.  Note however that $x_c$ has only a mild effect on the LP signal-to-noise ratio.  In practice, multiple values of $x_c$ would be used to select clusters in a shear-selected survey \citep[see][]{Schirmer2007}.  Since apertures with larger $x_c$ are more likely to introduce contamination from LPs, we choose $x_c = 2$ as our fiducial value from here on.

\begin{figure}
\begin{center}
\resizebox{8.0cm}{!}{\includegraphics{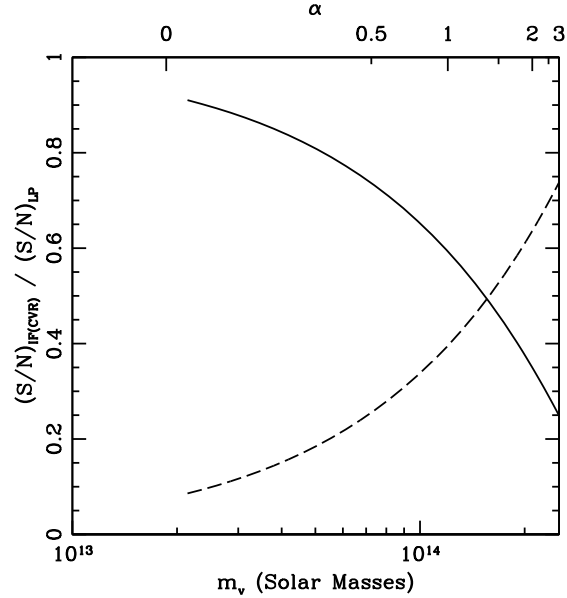}} 
\end{center}
\caption{Fractional contribution from the infall region (solid) and CVR (dashed) to the total LP signal-to-noise ratio as a function of virial mass $m_v$.  We assume a fixed total mass of $M = 10^{15}~\Msun$ and $z_l = 0.5$.  The power-law index $\alpha$ is varied so that the total LP signal-to-noise is 4 inside of an aperture with $\tout =  14$ arcminutes.  The top axis shows the $\alpha$ satisfying this criterion for each $m_v$.  Smaller virial masses result in a higher contribution from the infall region and flatter density profiles.}
 \label{StoNcontPL}
 \end{figure}

Figure \ref{StoNcontPL} shows the fractions $(S/N)_{\mathrm{IF}}/ (S/N)_{\mathrm{LP}}$ (solid) and $(S/N)_{\mathrm{CVR}}/ (S/N)_{\mathrm{LP}}$ (dashed) as a function of $m_v$ for $M = 10^{15}~\Msun$ and $z_l = 0.5$.  The power-law index $\alpha$ is varied to consistently satisfy $(S/N)_{\mathrm{LP}} = 4$ for $\tout = 14$.  The top axis shows the $\alpha$ required to satisfy this condition.  We truncate both curves at $\alpha = 3$ since the infall profile should not fall off more quickly than the NFW profile at large radii.  As expected, the contribution from the infall region is greatest for smaller $m_v$.  In this case, the density profile within the infall envelope must fall off slowly with radius in order to meet the signal-to-noise threshold.  On the other hand, larger virial masses result in a higher contribution from the CVR and steeper infall profiles.  The plot shows that the infall region makes a significant contribution to the total signal-to-noise, even up to $m_v \sim 1.5 \times 10^{14} ~\Msun$.
  
\section{Lensing Protoclusters as Dark Lenses} \label{LPsDL}

In the last section, we described a simple analytic model for LPs that allows us to compute the shear and aperture mass signal-to-noise ratio.  In what follows, we use the model to investigate the characteristics that a LP must possess in order to have the same observational signatures as a dark lens.

\subsection{X-ray luminosities and virial masses} \label{XRAYLUMINOSITIES}

It is well known that a cluster's X-ray luminosity scales with its virial mass.  Therefore, in order for a LP to be a plausible dark lens candidate, its virial mass must be low enough to avoid detection in relatively deep X-ray searches.  In this section, we use the semi-analytic calculation by \citet{Nord2008} to estimate the range of virial masses that a LP must have in order to be ``dark". 

\begin{figure}
\begin{center}
\resizebox{8.0cm}{!}{\includegraphics{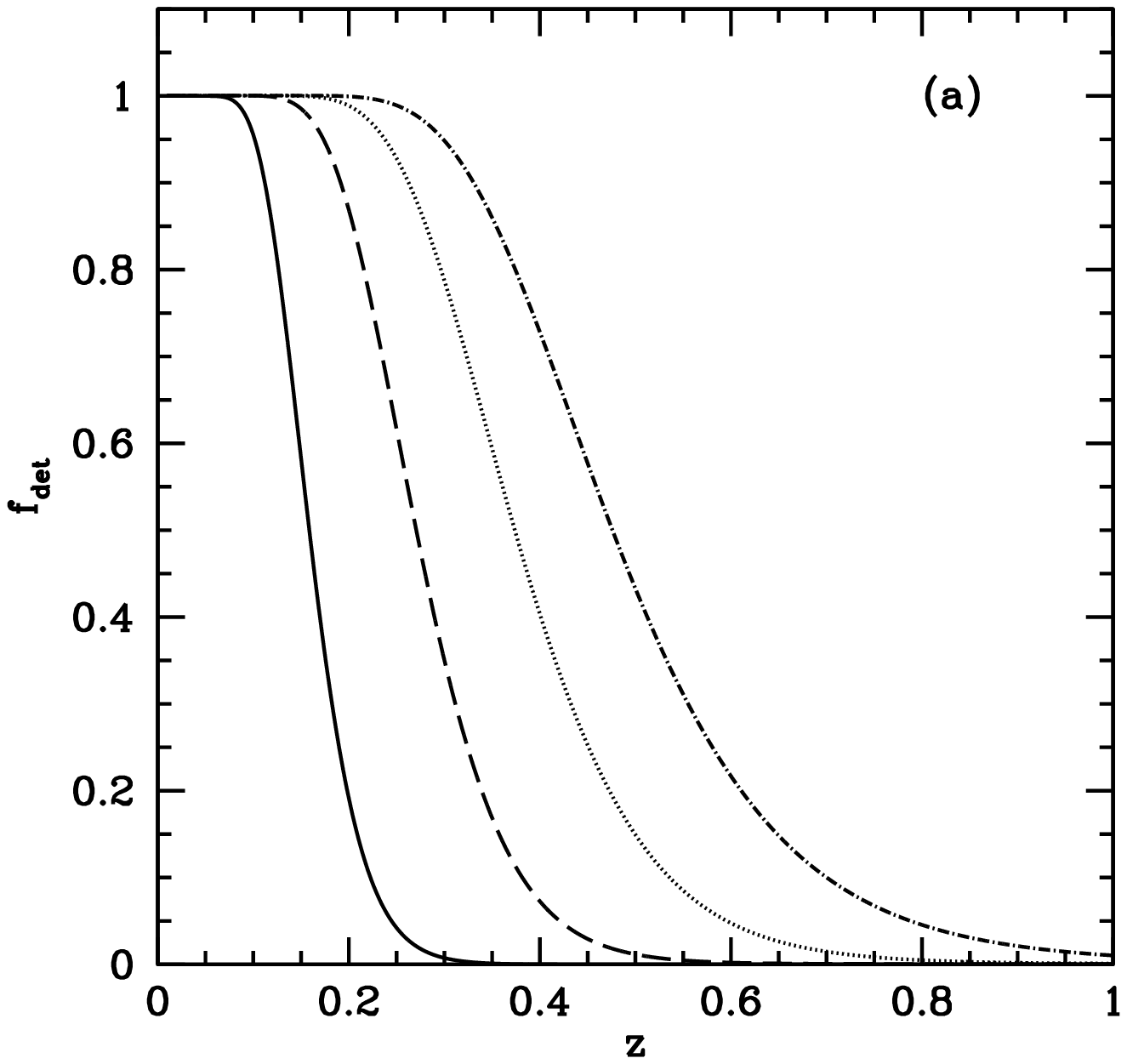}} \hspace{0.13cm}
\resizebox{8.0cm}{!}{\includegraphics{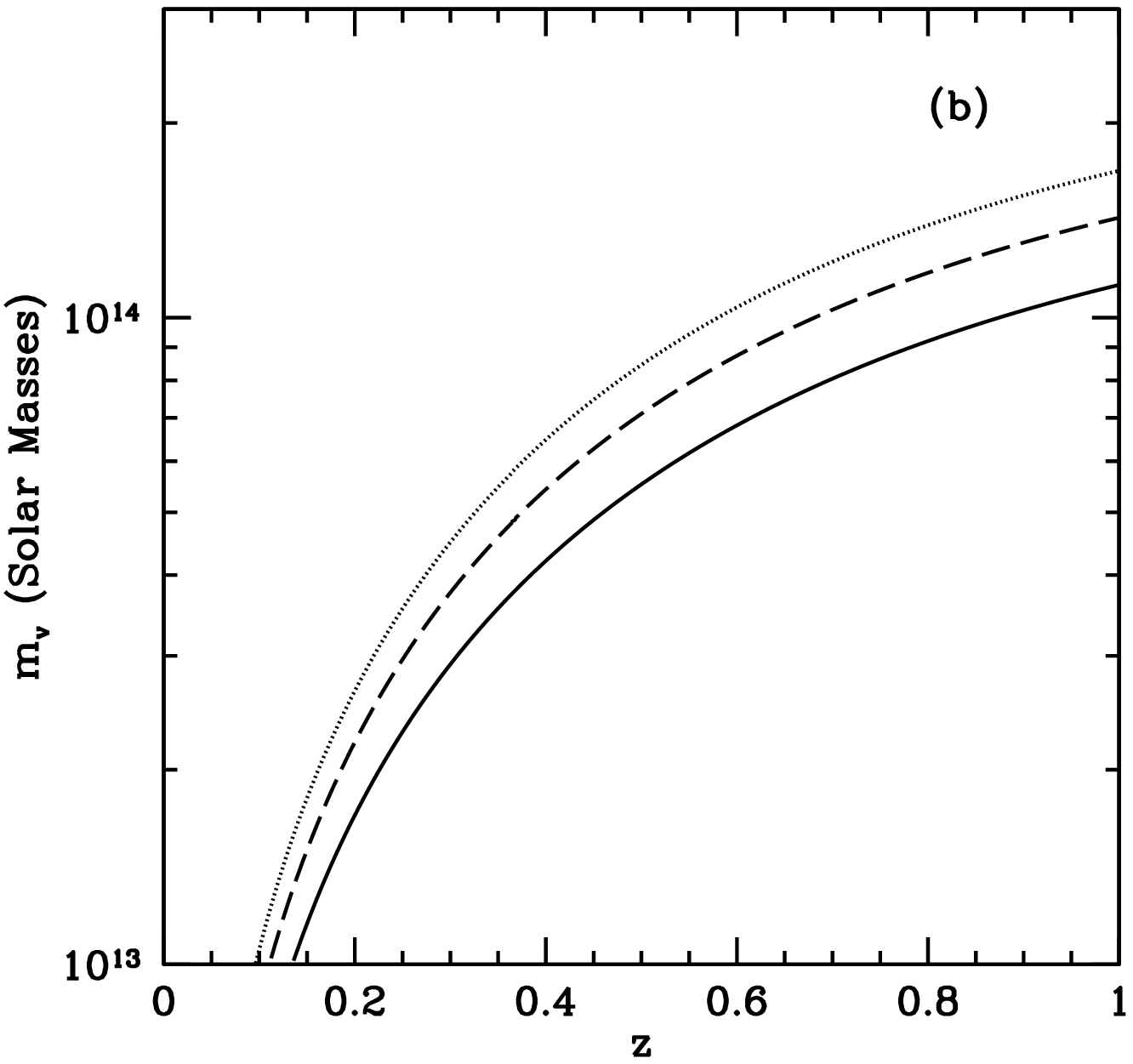}} 
\end{center}
\caption{Panel $(a)$:  the fraction of virialised haloes with a given mass detected via soft-band X-ray emission.  We assume a flux limit of $10^{-14}~\mathrm{ergs~s^{-1}~cm^{-2}}$ and luminosity dispersion of $\sigma_l = 0.59$. The solid, dashed, dotted, and dot-dashed curves correspond to halo masses of $m_v = 2$, $4$, $6$, and $8\times10^{13}~\Msun$ respectively.  Panel $(b)$:  the virial mass limit $m_v^d$ for a fixed detection fraction $f_{\mathrm{det}} = 0.1$ (solid), $0.3$ (dashed), and $0.5$ (dotted).  For a given redshift, less than $f_{\mathrm{det}}$ of haloes with mass $< m_v^d$ are detected.  }
 \label{XRayPL}
 \end{figure}

Given a sample's soft-band ($0.1- 2.4$keV) flux threshold, \citet{Nord2008} use the window function in equation (2) of their paper to model the fraction of virialised clusters detected (which we denote as $f_{\mathrm{det}}$) as a function of redshift.  Figure \ref{XRayPL} shows $f_{\mathrm{det}}$ for a flux limit of $10^{-14}~\mathrm{ergs~s^{-1}~cm^{-2}}$.  We assume a low flux limit here since a dark lens detection would likely be followed by a deep X-ray search.  The solid, dashed, dotted, and dot-dashed lines correspond to virial masses of $m_v = 2$, $4$, $6$, and $8~\times 10^{13}~\Msun$ respectively.  Following \citet{Nord2008}, we assume a luminosity dispersion of $\sigma_l = 0.59$.   At low redshifts ($z \la 0.05$), nearly $100 \%$ of virialised haloes with the above masses are detected as X-ray sources.  The percentage quickly declines with redshift.  At $z = 0.5$, roughly $0$, $1$, $15$, and $43$ per cent of virialised haloes are detected with masses of $2$, $4$, $6$, and $8~\times 10^{13}~\Msun$ respectively.  All detection fractions drop to nearly zero by $z = 1$.  

At a given redshift, we would ultimately like to estimate the maximum virial mass that a LP can have while still maintaining a small chance of being detected.  For this task, we assume a fixed detection fraction $f_{\mathrm{det}}$ and solve for the corresponding virial mass $m_v^d$ as a function of redshift.  Less than the fraction $f_{\mathrm{det}}$ of virialised haloes are detected below this virial mass limit.  The solid, dashed, and dotted lines in Figure \ref{XRayPL} $(b)$ show the virial mass limit $m_v^d$ for $f_{\mathrm{det}} = 0.1$, $0.3$, and $0.5$ respectively.  As an example, less than $10~\%$ of virialised haloes are detected below a mass of $m_v^d = 5.5\times10^{13}~\Msun$ at $z = 0.5$.  As expected, the plot shows that LP virial masses must be smaller at lower redshifts in order to maintain a significant chance of being undetected.  From here on, LPs with a low probability of being detected via soft-band X-ray emission will be called ``dark." 

Note that in the above calculation, we have assumed that soft-band X-ray luminosities of LPs roughly follow the mean scaling relation given by equation (1) of \citet{Nord2008}.  Determining the extent to which this assumption is valid is difficult due to the effects of accretion and mergers on a cluster's X-ray luminosity.  Using hydrodynamical simulations, \citet{Rowley2004} find that the accretion of sub-clumps creates scatter in the $L  - m_v$ relation by shifting clusters below the mean curve.  They attribute this to the fact that while both the mass and luminosity increase as a sub-clump falls toward the center of the cluster, the temperature typically stays constant or decreases slightly.  Unfortunately, the task of adequately addressing the above issue is beyond the scope of our simple analytic approach.  Further numerical studies are required to include these effects.

It should also be noted that our discussion is restricted to cases where the mass growth rate of a LP is dominated by the accretion of smaller sub-clumps (ie that the CVR is the largest progenitor halo).  In these cases where the sub-clumps are significantly cooler than the CVR, we expect the latter to be the dominant contributor to the integrated X-ray luminosity.  This assumption may not be true for cases where the infall region contains a group that is of similar mass to the CVR. However, these objects should be morphologically different from our model, and therefore represent a different class.

\subsection{Aperture mass detection} \label{APERTUREMASSDETECTION}
 
Armed with an appropriate range of dark LP virial masses, we turn our attention to shear selection. In this section we explore the physical characteristics of LPs that meet the aperture mass detection threshold of $S/N = 4$.
 
In order to compare the densities of shear-selected LPs to virialised clusters, we calculate the minimum LP over-density required to meet the $S/N$ threshold.  The $S/N$ of a LP is a function of $M$, $m_v$, $z_l$, and $\alpha$.  For a fixed $M$, $m_v$, and $z_l$, we solve for the power law index $\alpha$ such that $S/N = 4$ inside of a lens-centered aperture.  We then calculate the truncation radius $R$  and the average over-density, $\dNLlens$, using $1+\dNLlens = 3 M/(4 \pi \rhoBR R^3)$, where $\rhoBR$ is the mean matter density of the Universe. 

\begin{figure}
\begin{center}
\resizebox{5.7cm}{!}{\includegraphics{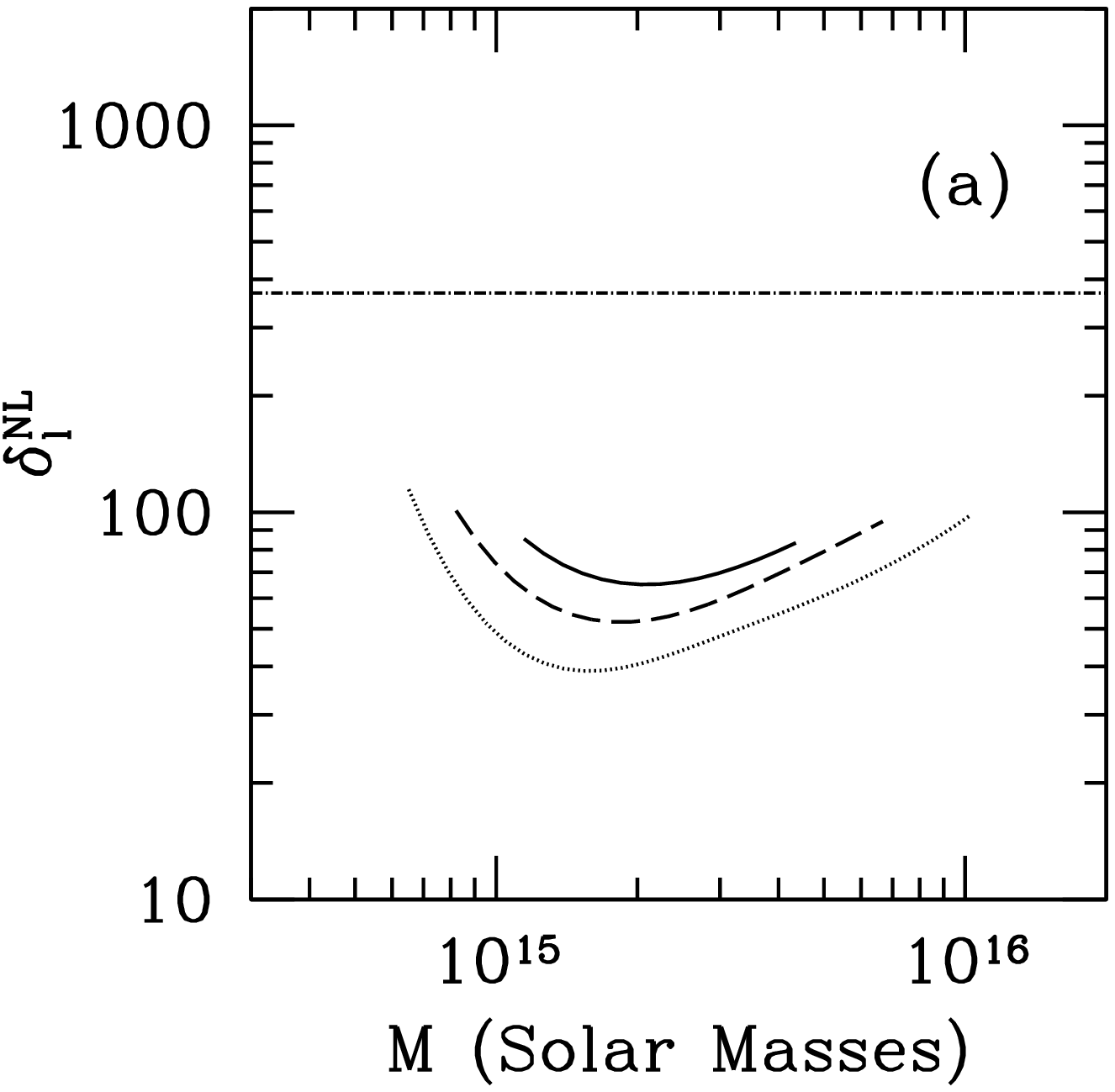}} \hspace{0.13cm}
\resizebox{5.7cm}{!}{\includegraphics{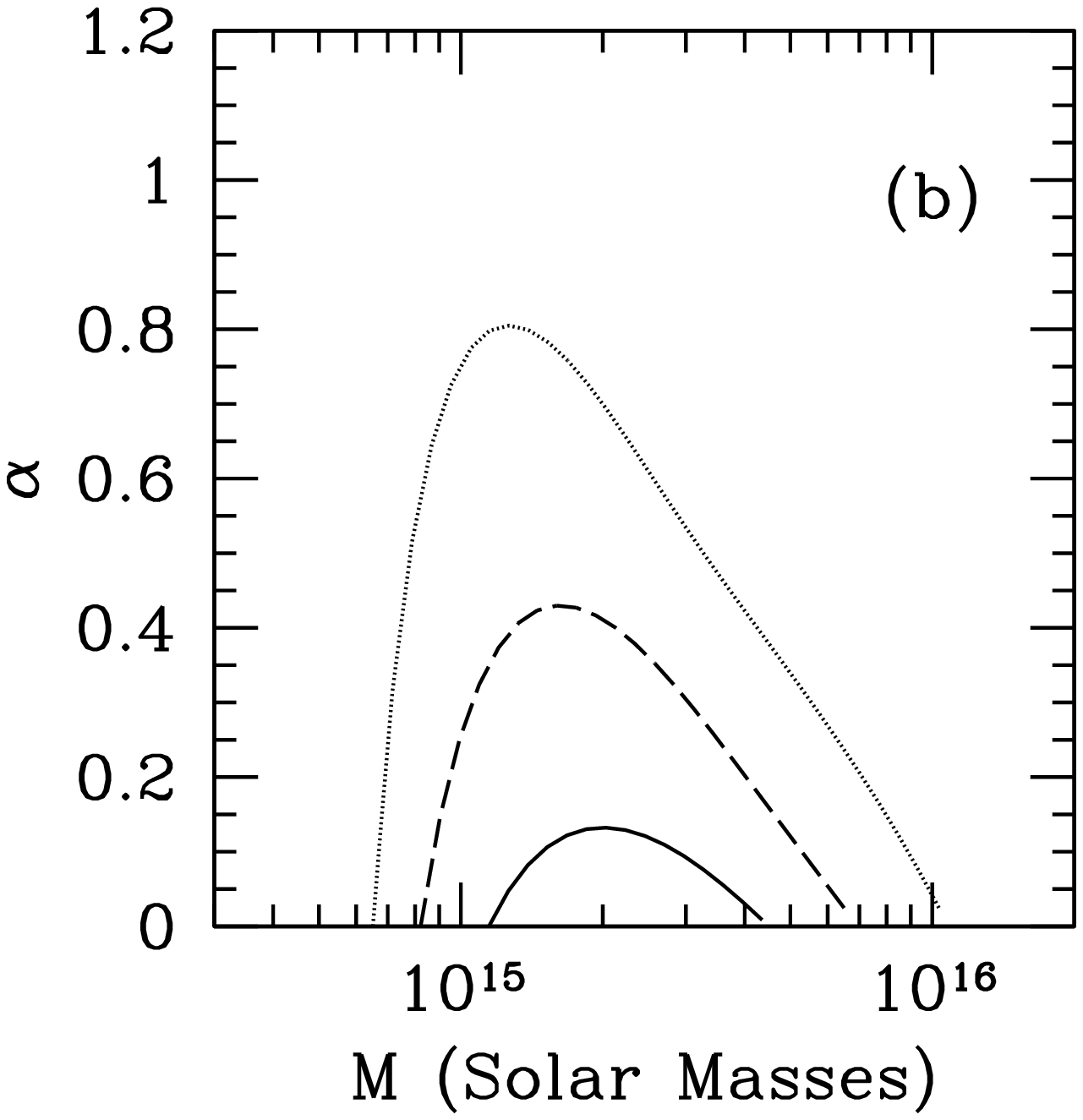}} \hspace{0.13cm}
\resizebox{5.7cm}{!}{\includegraphics{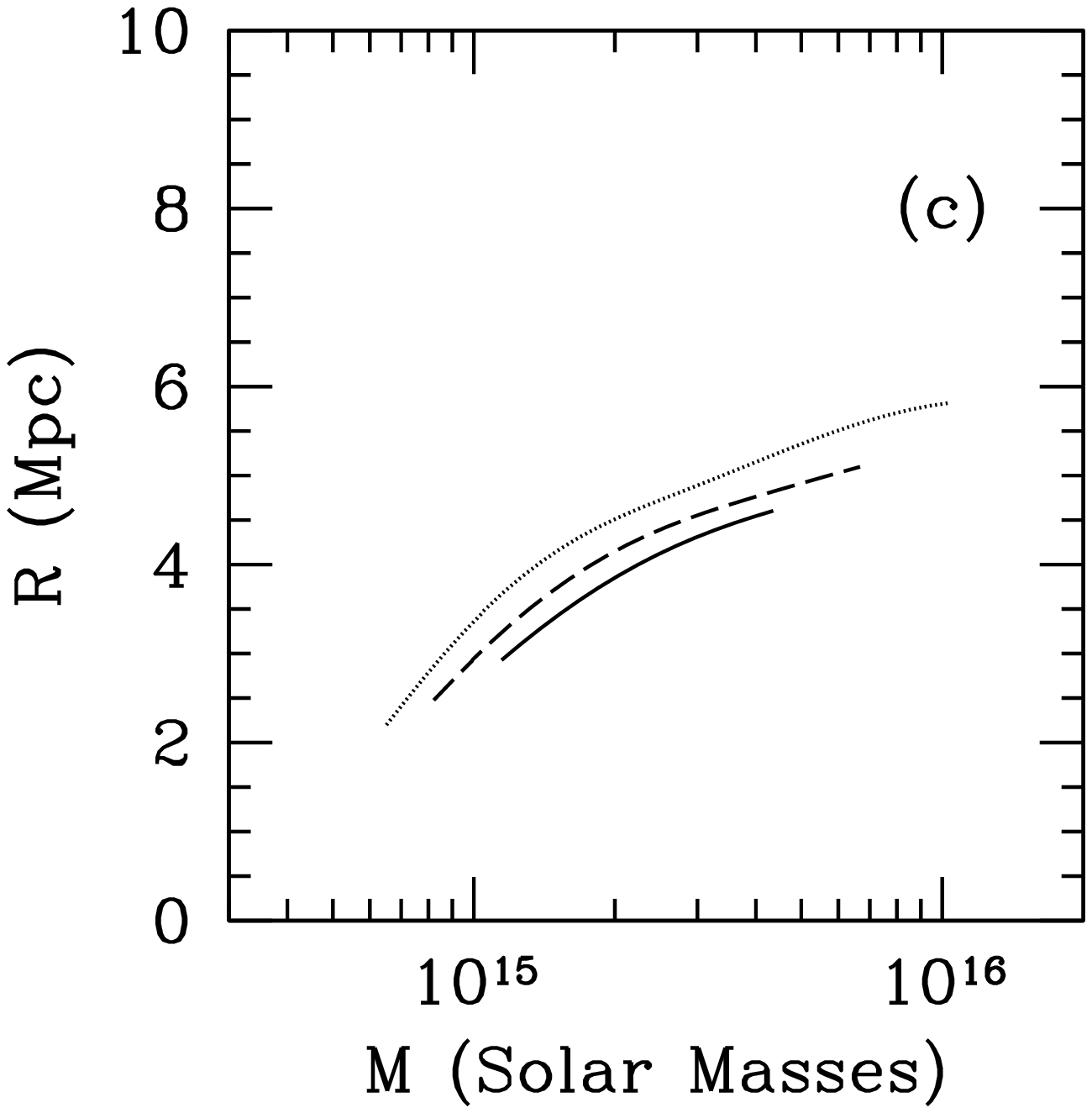}}
\end{center}
\caption{Panel $(a)$:  the average over-density required to produce an aperture mass signal-to-noise ratio of 4 inside of $R_{\mathrm{out}} = 5~\Mpc$ as a function of the total LP mass $M$.  The solid, dashed, and dotted lines correspond to $m_v = 10^{13}$, $5\times10^{13}$, and $10^{14}~\Msun$ respectively.  Also shown is the virialisation threshold at $z = 0.5$ (dot-dashed).  Panels $(b)$ and $(c)$ show the infall power-law index $\alpha$ and truncation radius $R$ required to satisfy the signal-to-noise condition.  The aperture mass technique is most sensitive to LPs with $R \sim R_{\mathrm{out}}$.}
 \label{dlensAPL}
 \end{figure}

\begin{figure}
\begin{center}
\resizebox{5.7cm}{!}{\includegraphics{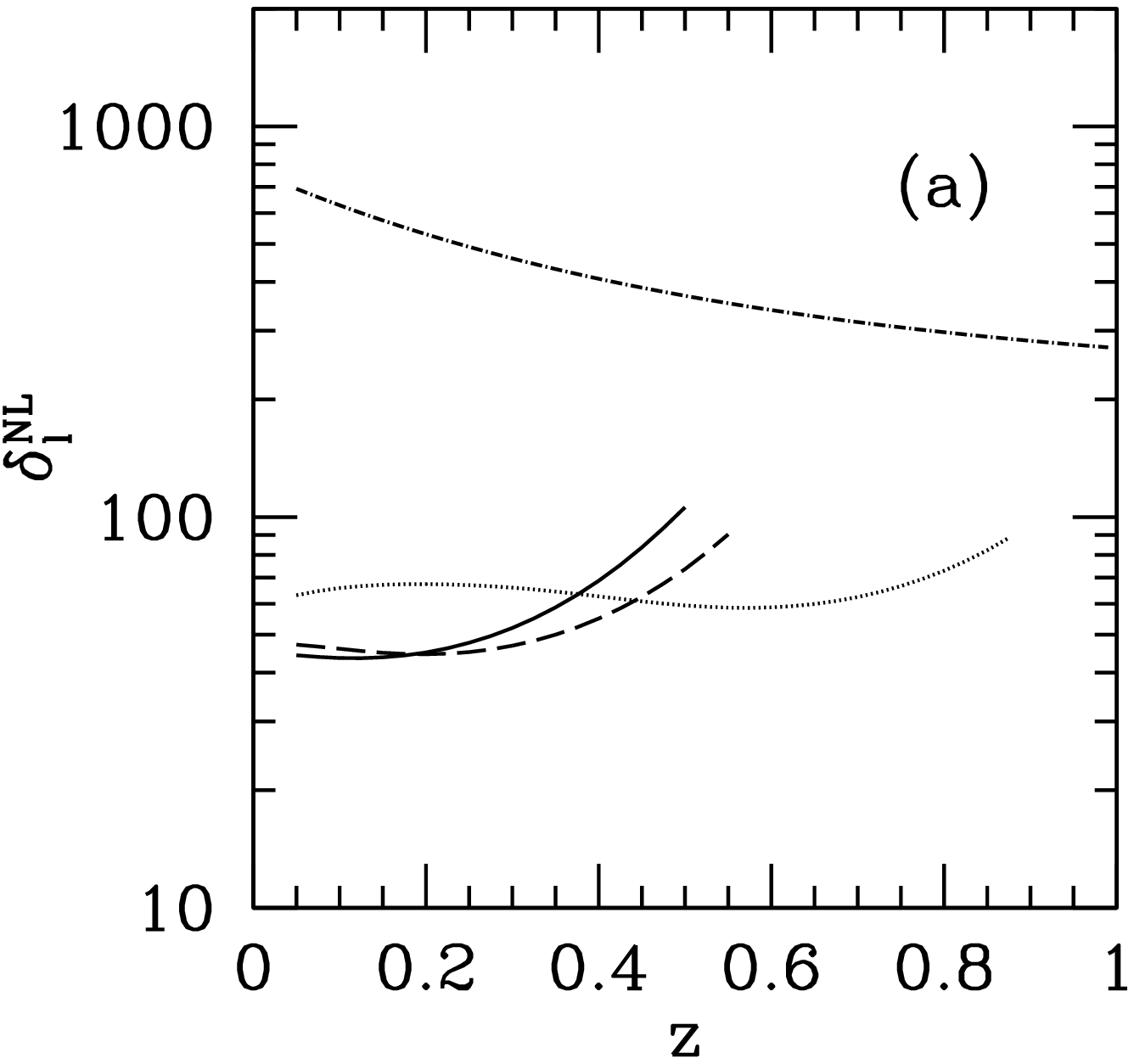}} \hspace{0.13cm}
\resizebox{5.7cm}{!}{\includegraphics{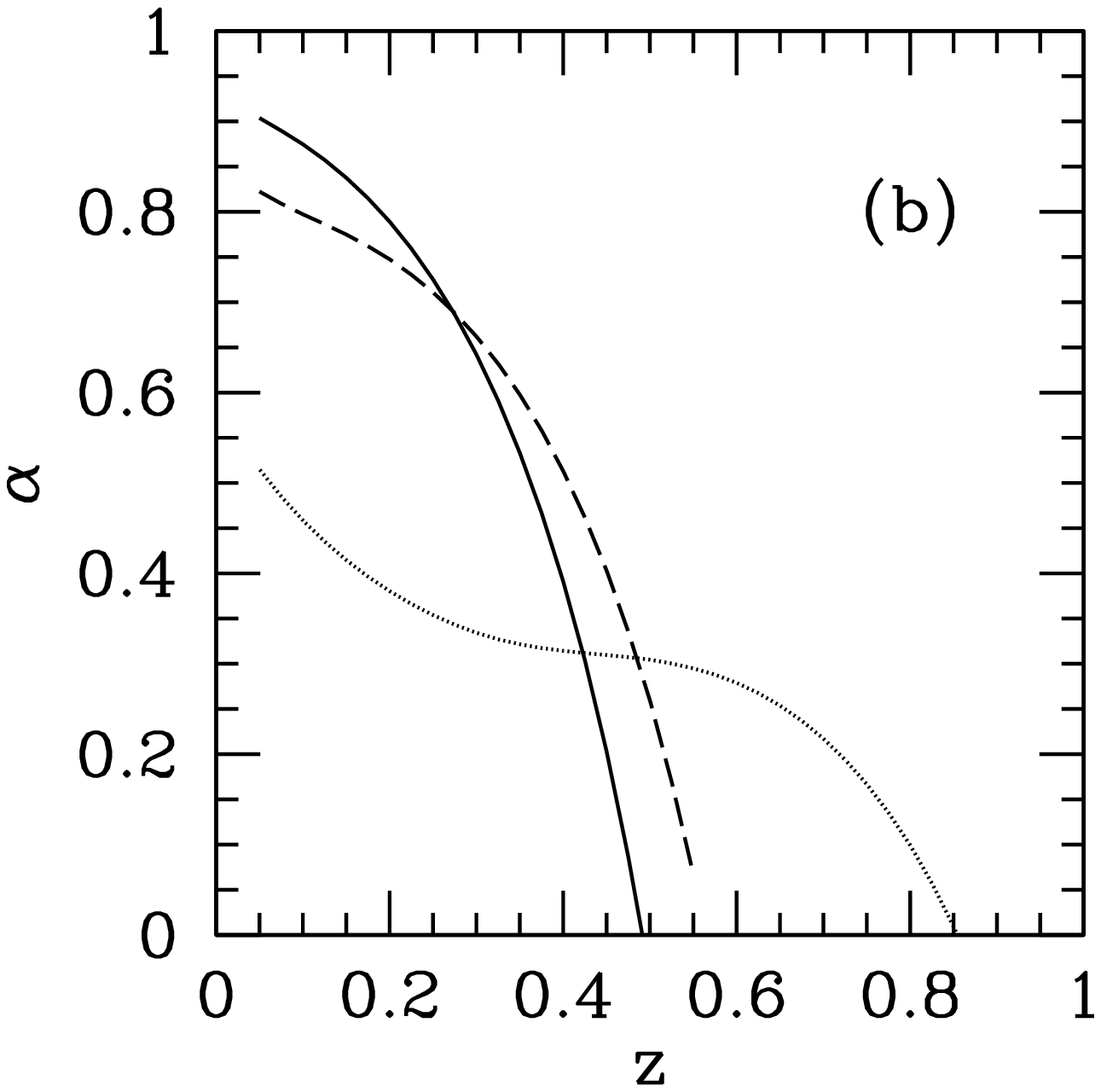}} \hspace{0.13cm}
\resizebox{5.7cm}{!}{\includegraphics{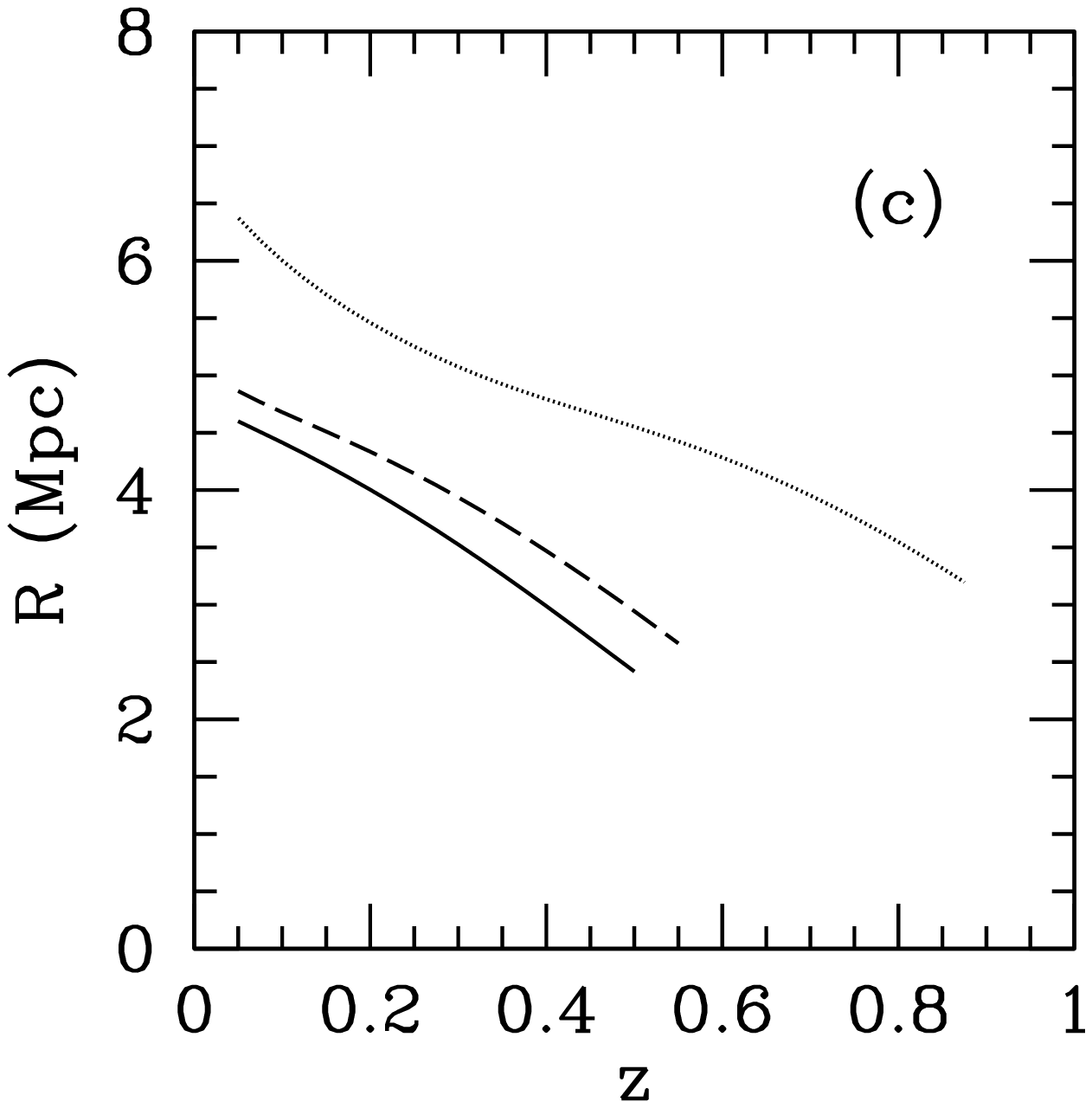}}\end{center}
\caption{Same as in Figure \ref{dlensAPL} shown as a function of redshift for a fixed virial mass of $m_v = 5\times10^{13}~\Msun$.  The solid, dashed, and dotted lines correspond to total masses of $M = 8\times10^{14}$, $10^{15}$, and $3\times10^{15}~\Msun$ respectively. }
 \label{dlensBPL}
 \end{figure}

The solid, dashed, and dotted curves in Figure \ref{dlensAPL} $(a)$ show $\dNLlens$ as a function of $M$ at $z_l = 0.5$ for $m_v = 10^{13}$, $5 \times10^{13}$, and $10^{14}~\Msun$ respectively.  These masses correspond to $f_{\mathrm{det}} \approx 0$, $0.05$, and $0.7$ at $z = 0.5$.  The dot-dashed line shows the virialisation threshold, $\delta^{\mathrm{NL}} = 200\rho_c/\rhoBR - 1 = 368$, at $z = 0.5$.  Panels $(b)$ and $(c)$ show the corresponding power-law index $\alpha$ and truncation radii $R$.  We assume a signal-to-noise threshold of $S/N = 4$ and aperture radius of $R_{\mathrm{out}} = 5~\Mpc$ for all curves.  Figure \ref{dlensBPL} shows $\dNLlens$, $\alpha$, and $R$ as a function of $z$ for $M = 8 \times 10^{14}$ (solid), $10^{15}$ (dashed) and $3 \times 10^{15}$ (dotted)$~\Msun$.  Here, we assume a fixed virial mass of $m_v = 5 \times 10^{13}~\Msun$.  

In panel $(a)$ of Figure \ref{dlensAPL} the minimum of $\dNLlens$ occurs when the truncation radius is similar to the aperture size.  The aperture mass measure is most sensitive to overdensities with scales close to the aperture radius.  The plots show that LPs must be increasingly over-dense to meet the $S/N = 4$ threshold as $M$ decreases beyond the minimum, since the amount of mass within the aperture decreases.  The over-density rises as $M$ increases beyond the minimum because, as more mass is added to the regions outside of the aperture radius, the integrated shear inside of $\tout$ decreases.  In other words, although $S/N$ increases at larger radii, the $S/N$ within $\tout$ actually decreases.  Hence, a higher over-density is required to produce $S/N = 4$ as the scale of lens exceeds the aperture radius.  

Figure \ref{dlensAPL} $(b)$ shows that LPs with a low probability of being detected via their X-Ray luminosities must have infall regions with flat power-laws in order to meet the aperture mass detection threshold.  We use the results of \citet{Tavio2008} to qualitatively determine whether such objects are common in their high resolution N-body simulations.  The fits to equation (12) of their paper represent the density profiles obtained by averaging over all haloes in a given mass bin.  For fixed parameters given in \citet{Tavio2008}, we calculate the effective $\alpha$ required to reproduce the mass enclosed by their profile inside of $10\rvir$.  Using a virial mass of $4.8\times 10^{13}~\Msun$ and concentration $6.82$, we calculate a value of $\alpha = 2.04$, indicating that the flat profiles shown in Figure \ref{dlensAPL} $(b)$ are likely to be rare.  In the next section, we will analytically estimate how rare these objects are using the excursion set formalism.  
 
The plots also show that there are minimum and maximum detectable LP masses, which we denote as $M_{\mathrm{min}}(m_v,z)$ and $M_{\mathrm{max}}(m_v,z)$ respectively. For a fixed virial mass and redshift, $M_{\mathrm{min}}(m_v,z)$ and $M_{\mathrm{max}}(m_v,z)$ are determined by where $\alpha = 0$ for constant $S/N  = 4$.  Using $\alpha = 0$ as the minimum allowed power-law index ensures that we never consider cases where the LP density profile increases with radius.  On the other hand, we could just as easily use some other non-zero minimum power-law index, $\alpha_{\mathrm{min}}$.  As Figure \ref{dlensAPL} $(b)$ shows, the effect of choosing some $\alpha_{\mathrm{min}} > 0$ is simply to reduce the interval of detectable $M$.  

Finally, we note that it is the signal-to-noise requirement that determines $\alpha$ and $R$; we impose no dynamical constraints on these parameters.  Therefore, not all of the values of $\alpha$ and $R$ shown above may be physical.  As an example, consider the dotted curve in Figure \ref{dlensAPL} $(a)$, corresponding to $m_v = 10^{14}~\Msun$.  It indicates that solutions exist for $M \sim 10^{16}~\Msun$ that have extremely flat ($\alpha \approx 0$) density profiles that extend out to $R \sim 6~\Mpc$.  

One way to eliminate some of the more extreme cases is to impose a higher value for $\alpha_{\mathrm{min}}$.  As Figures \ref{dlensAPL} $(b)$ and $(c)$ show, doing so both ensures that the density profiles fall off reasonably with $r$ and removes the cases with the largest truncation radii.  However, without any other dynamical arguments, choosing an appropriate $\alpha_{\mathrm{min}}$ is somewhat arbitrary.  Given the large variations observed by \citet{Tavio2008} beyond the virial radius, extremely flat logarithmic slopes are possible, though the abundance of such infall regions has yet to be investigated.  Since there is no obvious choice for $\alpha_{\mathrm{min}}$, we will display results for multiple values in section \ref{IMPLICATIONS}.   
 
\section{Abundances} \label{ABUNDANCES}

\subsection{The excursion set formalism} \label{EXCSET}

The excursion set formalism \citep{PandS1974,Bond1991,LaceyCole1993} was developed to infer the statistical properties of the non-linear density field using the framework of linear perturbation theory.  Perhaps the most well known example is the derivation of the halo mass function from linear theory, often referred to as the Press-Schecther (PS) mass function.  However, one great advantage of the excursion set approach is that its basic framework can be applied to a wide variety of abundance calculations.  Its versatility stems from the fact that, in principle, it may be applied to arbitrary linear over-density thresholds.  These thresholds are analogous to the critical density $\delta_c$ in the Press-Schechter halo abundance problem.  In what follows, we will describe how the formalism may be used to obtain the abundance of a general object $A$ defined by the linear over-density threshold $\delta_A$.  In section \ref{MASSFUNCTION}, we will fix $\delta_A$.

Consider a point $\mathbf{x}$ in space within a realization of the matter density field at any early epoch in the Universe, before the growth of perturbations enters the non-linear regime.  Rather than attempt to solve the non-linear evolution of density perturbations, we linearly extrapolate the initial density field to a later epoch using the growth factor from linear perturbation theory, $D(z)$.  The linear over-density at a point $\mathbf{x}$ at a later time $z$ is simply given by $\delta(z,\mathbf{x}) = \delta(z_i,\mathbf{x})D(z)/D(z_i)$, where $z_i$ is the initial redshift.  We then assume that the statistical properties of the true density field (for example, the halo abundance) at a given redshift can be inferred to a reasonable extent from the linearly extrapolated initial density field.

For convenience, it is common practice to linearly extrapolate the initial density field to the present day.  The over-density at a point $\mathbf{x}$ becomes redshift independent, but the over-density threshold becomes $\delta_A(z) = \delta_A/D(z)$, where we have normalized $D(z)$ to unity at the present day.  In what follows we adopt this convention.  Whenever it is necessary to discuss quantities that are not linearly extrapolated to the present day, we will note it in the text.  
 
The linear over-density around the point $\mathbf{x}$ is smoothed with a window function $W(r;R_W)$ of scale $R_W$ to obtain

\begin{equation}
\delta(R_W) = \int{d^3rW(r;R_W)\delta(r)}.
\end{equation}  
Owing to the mathematical simplicity it affords, the most common choice for the window function is the $k$-space top-hat window, defined by

\begin{equation}
W(k;R_W) =   \left\{ \begin{array}{ll} 1 & \mbox{($k \le R^{-1}_W$)} \\ 
	0 & \mbox{($k > R^{-1}_W$).} \end{array} \right.
\label{SharpKeq}
\end{equation}   
 
One starts by smoothing the density field around $\mathbf{x}$ for large $R_W$ (small k), and lowering $R_W$ in increments.  For each $R_W$, the variance of overdensities smoothed on this scale in an ensemble of density fields is calculated, 

\begin{equation}
S(R_W) \equiv \sigma^2(R_W) = \frac{1}{2 \pi^2}\int_0^{1/R_W}{dk~k^2 P(k)},
\label{Seq}
\end{equation}
where $P(k)$ is the linear power spectrum.  The set of points $\{S(R_W),\delta(R_W)\}$ traces out a trajectory parameterised by $R_W$ in the $\{S,\delta\}$-plane.  In the limit that $\Delta S \rightarrow 0$, \citet{Bond1991} showed that the probability density $Q(S,\delta)$ for a trajectory at $\{S,\delta\}$ satisfies

\begin{equation}
\frac{\partial Q}{\partial S} = \frac{1}{2}\frac{\partial^2 Q}{\partial \delta^2}.
\label{diffusioneq}
\end{equation}

We now turn to the case where there is an over-density threshold defining object $A$ as discussed above.  When $\delta(R_W)$ moves above or below $\delta_A$ (depending on the particular application) at a scale $S(R_W)$, the point $\mathbf{x}$ is assumed to be within an object of that scale.  The goal then is to calculate the fraction of trajectories that cross $\delta_A$ between the scales $S$ and $S+\dd S$.  Mathematically, this is realized by solving the diffusion equation (\ref{diffusioneq}) with absorbing barrier $\delta_A$.  When the boundary condition $Q(S,\delta_A) = 0$ and initial condition $Q(S_0,\delta) = \delta_{\mathrm{D}}(\delta - \delta_0)$, where $\delta_{\mathrm{D}}$ is the Dirac delta function, are applied to equation (\ref{diffusioneq}), $Q(S,\delta|S_0,\delta_0)~\dd \delta$ represents the probability that a trajectory starting at $\left\{ S_0,\delta_0 \right\}$ obtains an over-density between $\delta$ and $\delta + \dd \delta$ at $S$ without having crossed $\delta_A$.  The fraction of trajectories that cross the threshold at or prior to $S(R_W)$ is given by the complement of $Q$,

\begin{equation}
F(S,\delta_A|S_0,\delta_0) = 1 - \int_{-\infty}^{\delta_A}{Q(S,\delta|S_0,\delta_0)~d\delta}.
\label{massfractioneq}
\end{equation}
Equation (\ref{massfractioneq}), which applies only to the case where up-crossings are of interest, represents the fraction of mass within objects A with mass greater than $M(R_W)$.  Hence, the differential fraction of mass within objects $A$ is given by

\begin{equation}
f_S(S,\delta_A|S_0,\delta_0) = \frac{\dd F(S,\delta_A|S_0,\delta_0)}{\dd S} = -\frac{1}{2}\left[ \frac{\partial Q}{ \partial \delta} \right]^{\delta_A}_{-\infty}
\label{firstcrossingeq}
\end{equation}
where the last equality was obtained by using equation (\ref{diffusioneq}).  Equation (\ref{firstcrossingeq}) is often referred to as the first-crossing distribution.  Taking $\left\{ S_0,\delta_0 \right\} = \left\{ 0,0 \right\}$ (ie that the density field approaches the mean when smoothed on large scales), the mass function of object $A$ may be obtained from

\begin{equation}
n(M)~dM = \frac{\rhoBR}{M}\left| \frac{dF(S,\delta_A)}{dS} \right| \frac{dS}{dM}~dM.
\label{massfunceq}
\end{equation}

Note that up until this point, we have assumed the use of the $k$-space top-hat window function.  However, the main disadvantage of equation (\ref{SharpKeq}) is that both the volume and mass within $W(k;R_W)$ is not well defined.  To overcome these problems, it is common practice to derive equation (\ref{massfunceq}) using the $k$-space filter, but at the end replace the variance $S(k)$ by the real space top-hat relation,

\begin{equation}
S(R_W) = \frac{1}{2\pi^2}\int{dk~k^2P(k) \left[ \frac{3\sin(kR_W) - 3kR_W\cos(kR_w)}{(kR_W)^3} \right]^2}.
\end{equation}  
In this case, the mass within the \emph{Lagrangian} radius $R_W$ is given by $M = \rhoBR~4\pi R_W^3/3$.  Therefore, in the final relations that we derive, we are to interpret the quantity $\sigma^2$ as the variance of the smoothed linear over-density inside of the Lagrangian radius $R_W$.  Similarly, the smoothed over-density $\delta$ is to be interpreted as the average over-density inside of $R_W$.

\subsection{The mass function of dark LPs} \label{MASSFUNCTION}

\begin{figure}
\begin{center}
\resizebox{8.0cm}{!}{\includegraphics{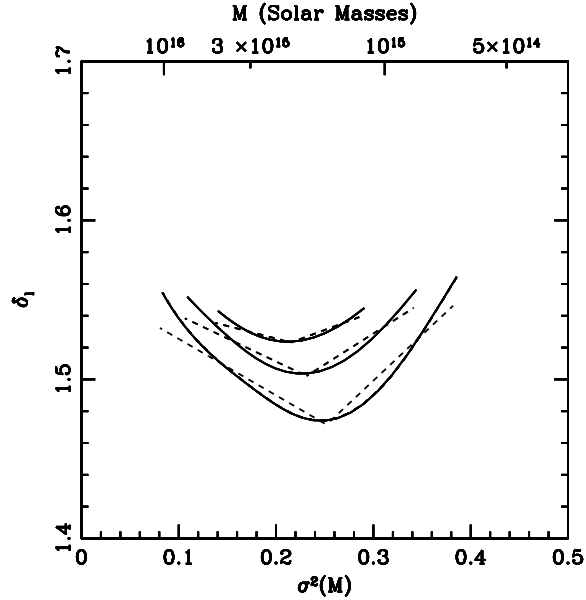}} 
\end{center}
\caption{The linear over-density of a LP required to create $S/N = 4$ inside of $R_{\mathrm{out}} = 5~\Mpc$ versus the variance of the linearly extrapolated density field, $\sigma^2(M)$.  From top to bottom, the solid curves show the exact linear over-density for $m_v = 10^{13}$, $5\times10^{13}$, and $10^{14}~\Msun$ respectively.  The dashed lines show the corresponding piecewise approximations (see appendix \ref{APPENDIXA}).  Note that the ordinate is not linearly extrapolated to the present day.}
 \label{lindlensPL}
\end{figure}  

In this section, we aim to estimate the abundance of dark LPs using the excursion set formalism  Unfortunately, the formalism cannot be used to calculate the abundance of objects with density profile (\ref{LPprofileEQ}).  Instead, we content ourselves with the more modest goal of considering overdensities that are large enough to create a weak lensing signal, but are unlikely to be detected via their X-Ray luminosities.  Our approach contains two steps: Step A) we add up the fraction of mass within overdensities above a lensing threshold.  For this we use the overdensity barriers obtained in section \ref{APERTUREMASSDETECTION}.  Step B) we multiply by the fraction of these overdensities that have a low probability of being detected by their X-Ray luminosities.

We begin our discussion with Step A.  Since the excursion set formalism is based on linear perturbation theory, our first task is to convert the $\dNLlens$ barriers of section \ref{APERTUREMASSDETECTION} to linear overdensities, $\delta_l$, using the spherical collapse model.  Figure \ref{lindlensPL} shows the results of this conversion as a function of the scale $S = \sigma^2(M)$.  $M$ is shown on the top axis (Recall that $S$ is a monotonically decreasing function of $M$).  Here we assume $z_l = 0.5$  and $\alpha_{\mathrm{min}} = 0$.  From top to bottom, the solid curves correspond to the exact $\delta_l$ for $m_v = 10^{13}$, $5\times10^{13}$, and  $10^{14} ~\Msun$ respectively.  The dashed curves show the corresponding piecewise approximations discussed below and in appendix \ref{APPENDIXA}.  

Note that each barrier shown in Figure \ref{lindlensPL} is scale-dependent and cannot be expressed as an analytic function of $S$.  Obtaining exact analytic solutions for their first crossing distributions is therefore impossible.  For a fixed $m_v$, we address this issue by approximating the barrier as two lines, with the form of equation (\ref{barrierEQ}).  In appendix \ref{APPENDIXA}, we show that a solution for the first crossing distribution with an absorbing barrier of this form can be reduced to quadrature.  In what follows, we will use the approximate first crossing distribution $f_S(S,\delta_l)$, given by equation \ref{fcEQ}.  

We make two additional notes about the over-density barriers in Figure \ref{lindlensPL}.  First, $\delta_l$ is a relatively weak function of $m_v$, particularly for smaller values of $m_v$.  In fact, below $m_v = 10^{13}~\Msun$ the amplitude of $\delta_l$ is virtually independent of $m_v$.  Secondly, for two virial masses $m_v^a$ and $m_v^b$, where $m_v^a > m_v^b$, any trajectory that crosses $\delta_l(m_v^b)$ must also cross $\delta_l(m_v^a)$ at a larger total mass scale.  We therefore assume that $f_{S}\left[S,\delta_l(m_v) \right]$ yields the approximate fraction of mass in lensing overdensities that \emph{potentially} correspond to dark LPs with virial mass less than $m_v$.  However, many of the lensing overdensities satisfying $\delta>\delta_l$ contain large sub-haloes.  Following section \ref{XRAYLUMINOSITIES}, if the sub-halo masses are large enough, then they have a high probability of being detected via their X-Ray luminosities.  Hence, their host overdensities would not satisfy the ``dark" criterion. Our goal, then, is to calculate the fraction of  overdensities whose sub-haloes do not exceed a maximum probability $f_{\mathrm{det}}$ of being detected via X-Ray luminosity (Step B).  Using the results of section \ref{XRAYLUMINOSITIES}, we can map this probability to a maximum allowed sub-halo mass, $m_v^d$.  Put in another way, our goal is to obtain the fraction of mass contained in trajectories without ``nearby" trajectories that cross the virialisation threshold at mass scales greater than $m_v^d$. 

In what follows, we adopt the notation $S_v = \sigma^2(m_v)$.  We also denote the \emph{linear} virialisation threshold as $\delta_v$, which is obtained by applying the spherical collapse model to $ \delta^{\mathrm{NL}} = 200\rho_c(z)/\rhoBR(z) - 1$.    Using N-Body simulations \citet{CasasMiranda2002} find that the probability, $P_V(N_h,m | M,\delta)$, of having $N_h$ sub-haloes with mass greater than $m$ within an over-density $\delta$ of mass $M$ is well described by a Gaussian with mean number

\begin{equation}
\left< N \right>(m | M, \delta) = \int_{m}^{M}{\dd m_v~N(m_v,\delta_v |M, \delta)}
\label{MeanEQ}
\end{equation}
and variance

\begin{equation}
\mathrm{Var}(N_h) = \left(1+A~D^2(z)S\right)\int_{m}^{M}{ \int_{m}^{M-m_1}{\dd m_1 \dd m_2}N(m_1,\delta_v | M,\delta)~N(m_2,\delta_v | M - m_1,\delta')} + \left< N \right> - \left< N\right>^2,
\label{varEQ}
\end{equation}
where
\begin{equation}
\delta' = \delta_v - \frac{(\delta_v-\delta)}{1- (m_1/M)}
\end{equation}
and

\begin{equation}
N(m_v,\delta_v | M, \delta)~\dd m_v \equiv \frac{dS_v}{\dd m_v}~\frac{M}{m_v}f_{S_v}(S_v, \delta_v | S, \delta)~\dd m_v
\end{equation}
is the average number of virialised sub-haloes with mass between $m_v$ and $m_v + \dd m_v$.  The second term in the prefactor in equation (\ref{varEQ}) is a phenomenological term accounting for clustering effects.  For the constant $A$, we use $0.05$, which was calibrated to simulations by \citet{CasasMiranda2002}.

The probability that a $\delta_l$ over-density has zero sub-haloes in the mass range $\left( m_v^d, M \right)$ is given by $P_V(0 , m_v^d | M , \delta_l)$.  Therefore, the fraction of mass within dark LPs whose sub-haloes have a probability $\leq f_{\mathrm{det}}$ of displaying detectable X-ray emission is $f_S\left[S,\delta_l(m_v^d) \right]\times P_V(0,m_v^d | M , \delta_l)$, and the mass function is given by

\begin{equation}
n_D(M,z)= \frac{\rhoBR}{M}~\left| \frac{\dd S}{\dd  M}\right|~
f_S\left[S,\delta_l(m_v^d) \right]\times P_V(0,m_v^d | M , \delta_l)
\label{massfuncEQ}
\end{equation}

\begin{figure}
\begin{center}
\resizebox{5.7cm}{!}{\includegraphics{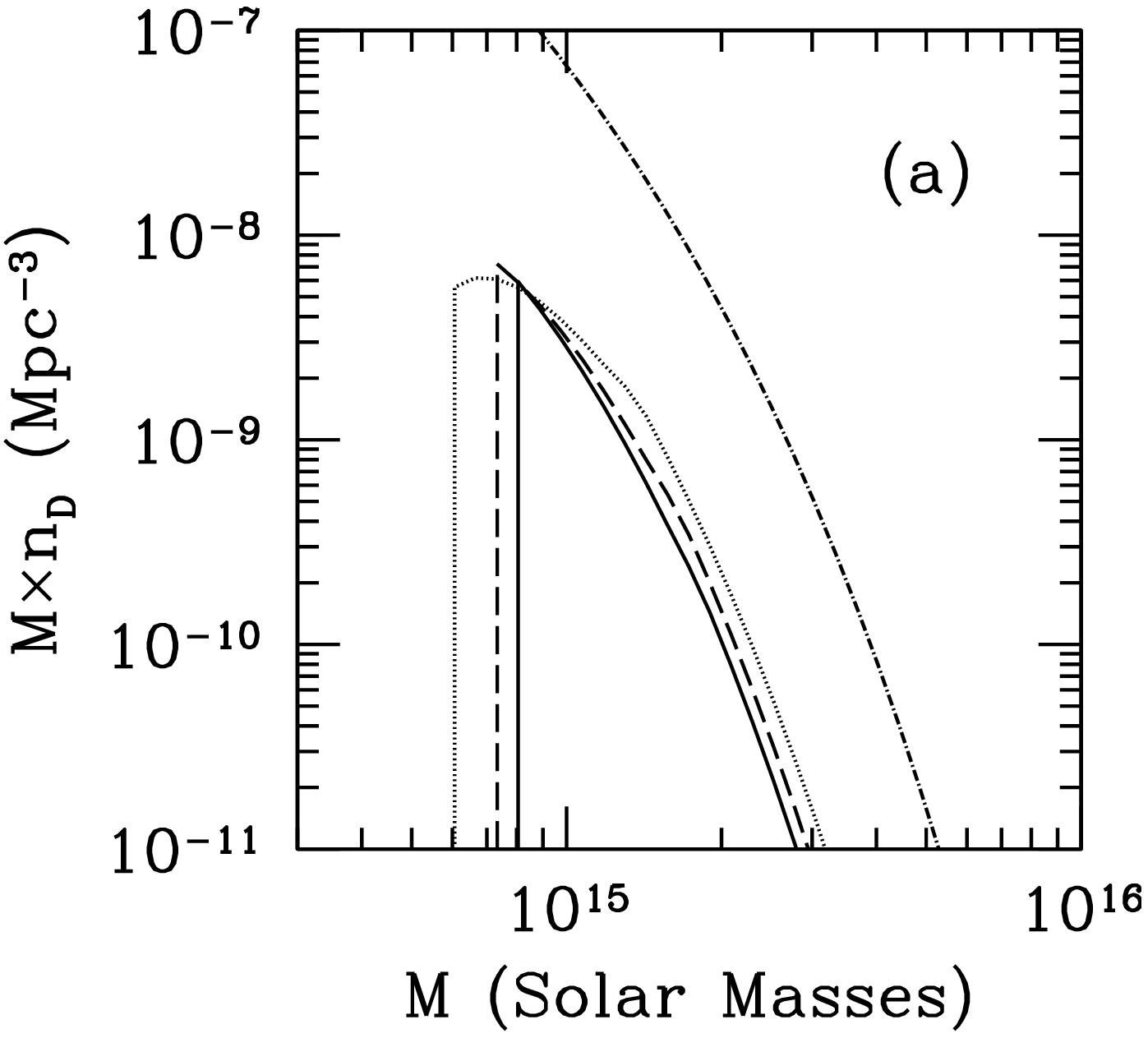}} \hspace{0.13cm}
\resizebox{5.7cm}{!}{\includegraphics{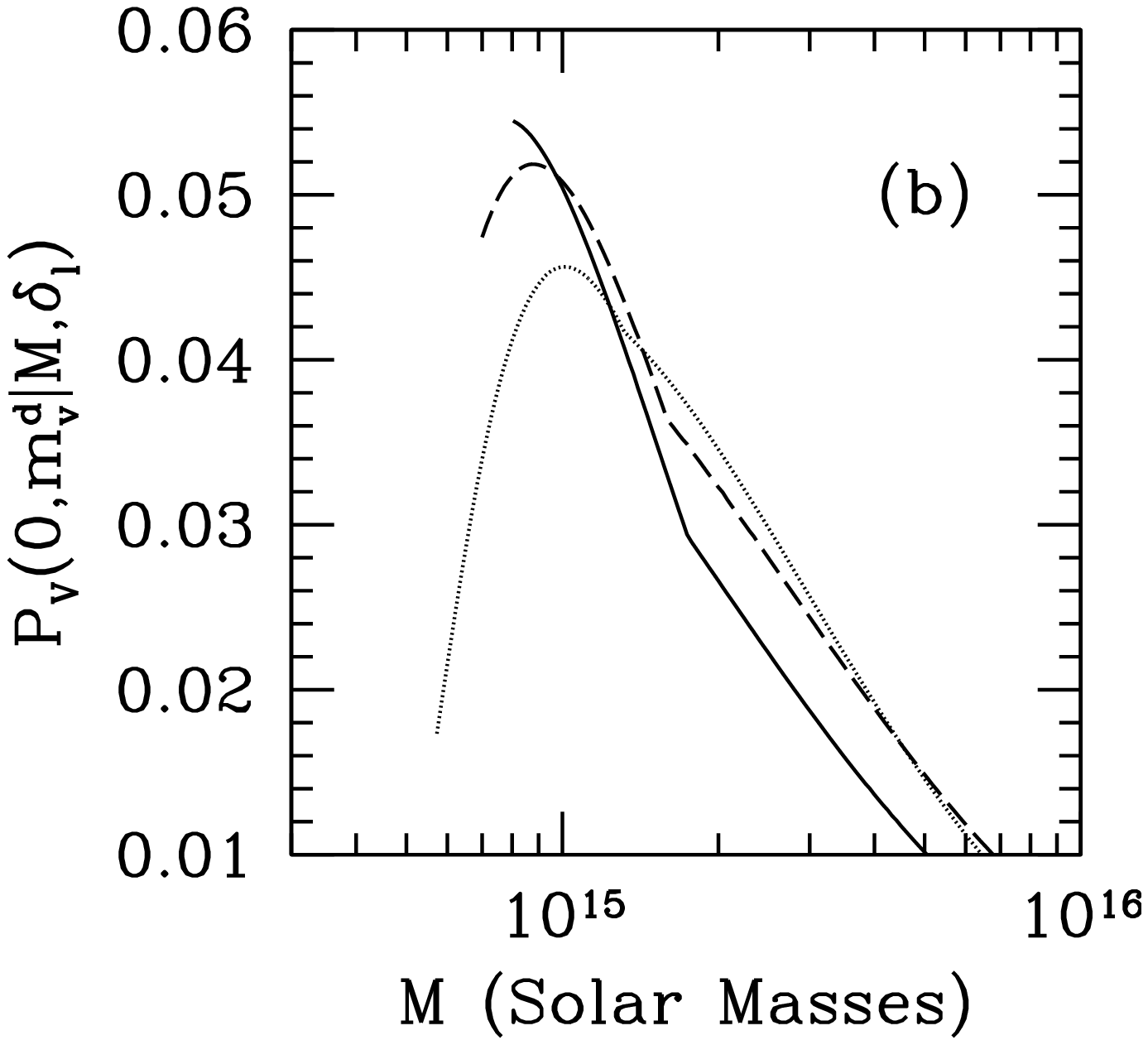}}
 \hspace{0.13cm}
\resizebox{5.7cm}{!}{\includegraphics{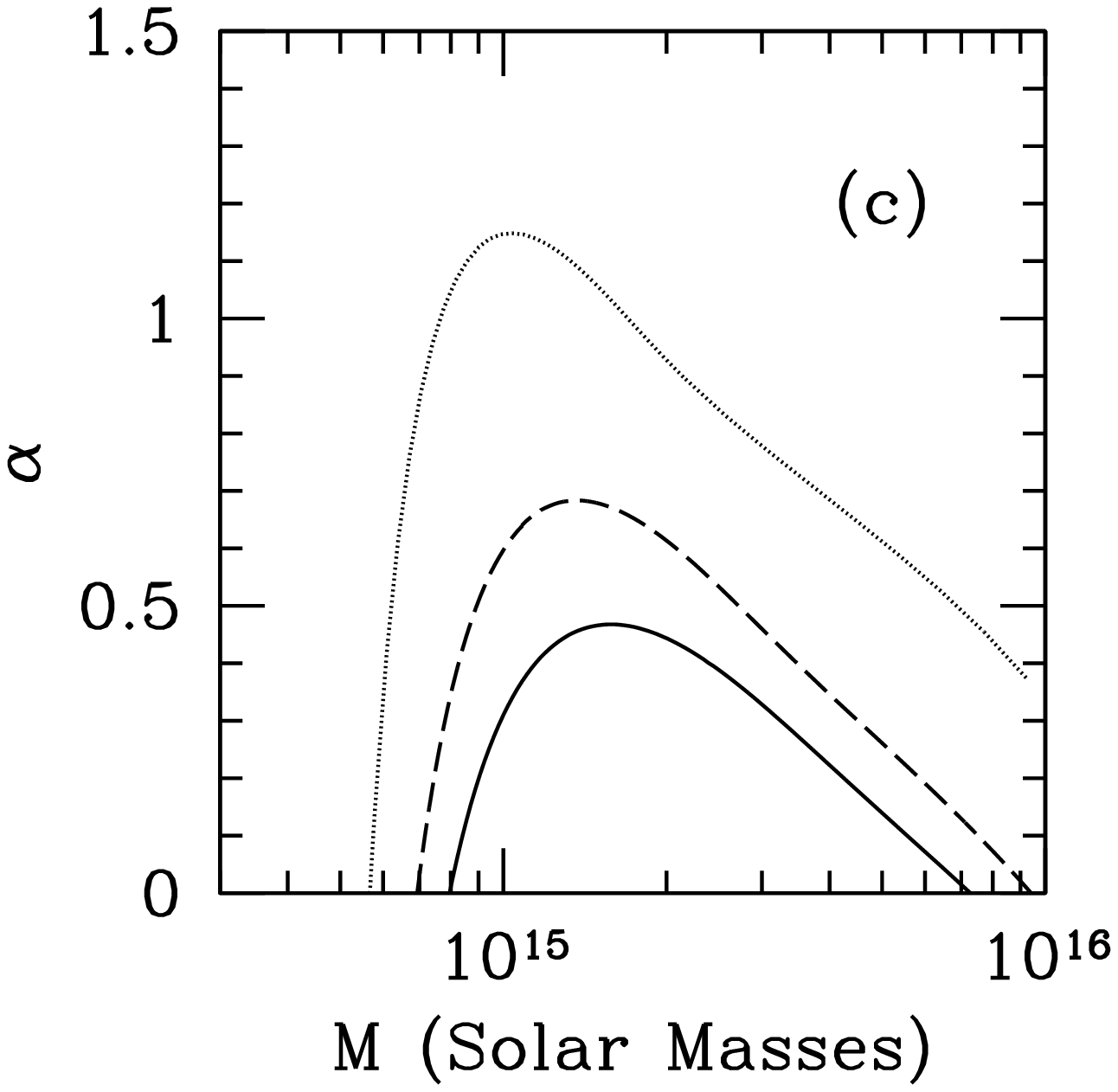}}
\end{center}
\caption{Panel $(a)$:  the dark LP mass function at $z_l = 0.5$. The solid and dashed lines correspond to $f_{\mathrm{det}} = 0.1$ ($m_v^d = 5.5\times10^{13}~\Msun$) and $0.5$ ($8.5\times10^{13}~\Msun$) respectively.  A soft-band X-Ray flux limit of $10^{-14}~\mathrm{ergs~s^{-1}~cm^{-2}}$ is assumed.  The dotted line corresponds to $f_{\mathrm{det}} = 0.1$ and a flux limit of $5\times10^{-14}~\mathrm{ergs~s^{-1}~cm^{-2}}$ ($m_v^d = 1.4 \times 10^{14}~\Msun$).  We assume $\alpha_{\mathrm{min}} = 0$ for all curves.  The dot-dashed curve shows the Sheth-Tormen mass function at the same redshift.  Panel $(b)$: the probability that a $\delta_l$ over-density has zero sub-haloes with mass between $m_v^d$ and $M$, shown for the same $m_v^d$ values above.  Since this probability is low in the mass range under consideration, the abundance of dark LPs is suppressed.  Panel $(c)$:  the infall power-law index required to satisfy $S/N = 4$ for $m_v = m_v^d$.  This panel serves as a reference for the types of LP profiles required to create a weak lensing signal in the mass range shown.}
 \label{darkLPMassFuncPL}
 \end{figure}

The solid and dashed lines in Figure \ref{darkLPMassFuncPL} $(a)$ show the dark LP mass function at $z_l = 0.5$ for $f_{\mathrm{det}} = 0.1$ and $0.5$, corresponding to sub-halo mass limits of $m_v^d = 5.5$ and $8.5 \times 10^{13}~\Msun$ respectively.  We assume the fiducial soft-band flux limit of $10^{-14}~\mathrm{ergs~s^{-1}~cm^{-2}}$.  The dotted line shows the mass function for $f_{\mathrm{det}} = 0.1$ and a flux limit of $5\times10^{-14}~\mathrm{ergs~s^{-1}~cm^{-2}}$ ($m_v^d = 1.4\times10^{14}~\Msun$).  A minimum allowed power law index of $\alpha_{\mathrm{min}} = 0$ is assumed for all curves.  The only notable effect of changing $\alpha_{\mathrm{min}}$ is to change the domain of detectable dark LP masses. As figures \ref{dlensAPL} $(a)$ and $(b)$ show, increasing $\alpha_{\mathrm{min}}$ truncates the low and high mass ends of detectable overdensities.  For reference, we show the \citet{SandT1999} (ST) halo mass function (dot-dashed) for $z = 0.5$.  

Panel $(b)$ shows $P_V\left[0,m_v^d | M , \delta_l(m_v^d) \right]$ for the same values of $m_v^d$.  The mass function of dark LPs is suppressed because the probability of finding a $\delta_l$ over-density without sub-haloes large enough for X-ray detection is small.  In addition, the large-mass end is suppressed in two ways:  1)   The shape of the over-density barriers in Figure \ref{lindlensPL} make it less likely for trajectories to cross at higher mass scales.  2)  Panel $(b)$ shows that as the total mass $M$ increases, it becomes less likely that the over-density will contain zero sub-haloes in the mass interval $(m_v^d,M)$.   Panels $(a)$ and $(b)$ also show that increasing $m_v^d$ through either $f_{\mathrm{det}}$ or the flux limit yields only a mild increase in the amplitude of the mass function.  Up to $m_v^d \sim M/2$, the probability factor $P_V\left[0,m_v^d | M , \delta_l(m_v^d) \right]$ remains low for increasing $m_v^d$ due to the decreasing variance (\ref{varEQ}).

Panel $(c)$ of Figure \ref{darkLPMassFuncPL} shows the power law index $\alpha$ required to create $S/N = 4$ for the LP profile.  The purpose of panel $(c)$ is to provide a reference for the types of density profiles required to create a detectable weak lensing signal in this mass range.  Note, however, that there is no direct relationship between $\alpha$ and the overdensities counted using the excursion set procedure in this section.  This is a limitation of analytic approach taken here.  The excursion set formalism does not yield information on the mass profiles of $\delta_l$ overdensities.  It is therefore impossible to rigorously quantify the aperture mass $S/N$ within the formalism.  However, we argue that by selecting objects with an adequate over-density to create a weak lensing signal, and correct sub-halo structure to avoid X-ray detection, we obtain a reasonable estimation of dark LP abundances. 

Finally, we note that we have not taken into account \emph{all} of the trajectories that may correspond to dark LPs.  Some trajectories may obtain $\delta > \delta_l$ at a scale $S = S_1$ (see appendix \ref{APPENDIXA}), corresponding to the maximum detectable LP mass $M_{\mathrm{max}}$.  A fraction $P_V$ of these trajectories correspond to dark LPs with mass $M_{\mathrm{max}}$.  However, since we have shown that large-mass LPs are extremely rare, we can neglect these trajectories with little consequence.   
 
\subsection{Dark LP counts and weak lensing surveys} \label{IMPLICATIONS}

Here we investigate the abundance of dark LPs and its consequences for future shear-selected cluster surveys.  For a fixed $f_{\mathrm{det}}$ the number counts of detectable dark LPs  per unit steradian, per unit redshift interval, is given by

\begin{equation}
\frac{\dd N_D(z)}{\dd z \dd \Omega} =
\frac{\dd V}{\dd z \dd \Omega} \int_{M_{\mathrm{min}}(z)}^{M_{\mathrm{max}}(z)}{n_D(M,z)~\dd M}
\label{darkLPnumbercountsEQ}
\end{equation}   
where $n_D$ is given by equation (\ref{massfuncEQ}).  Here,

\begin{equation}
\frac{\dd V}{\dd z \dd \Omega}  = \frac{c}{H_0}\frac{(1+z)^2 D_A(z)^2}{\sqrt{\Omega_m (1+z)^3 + \Omega_{\Lambda}}}
 \end{equation}
 is the comoving volume element, where $c$ is the speed of light and $H_0$ is the present-day Hubble parameter.
 
\begin{figure}
\begin{center}
\resizebox{5.7cm}{!}{\includegraphics{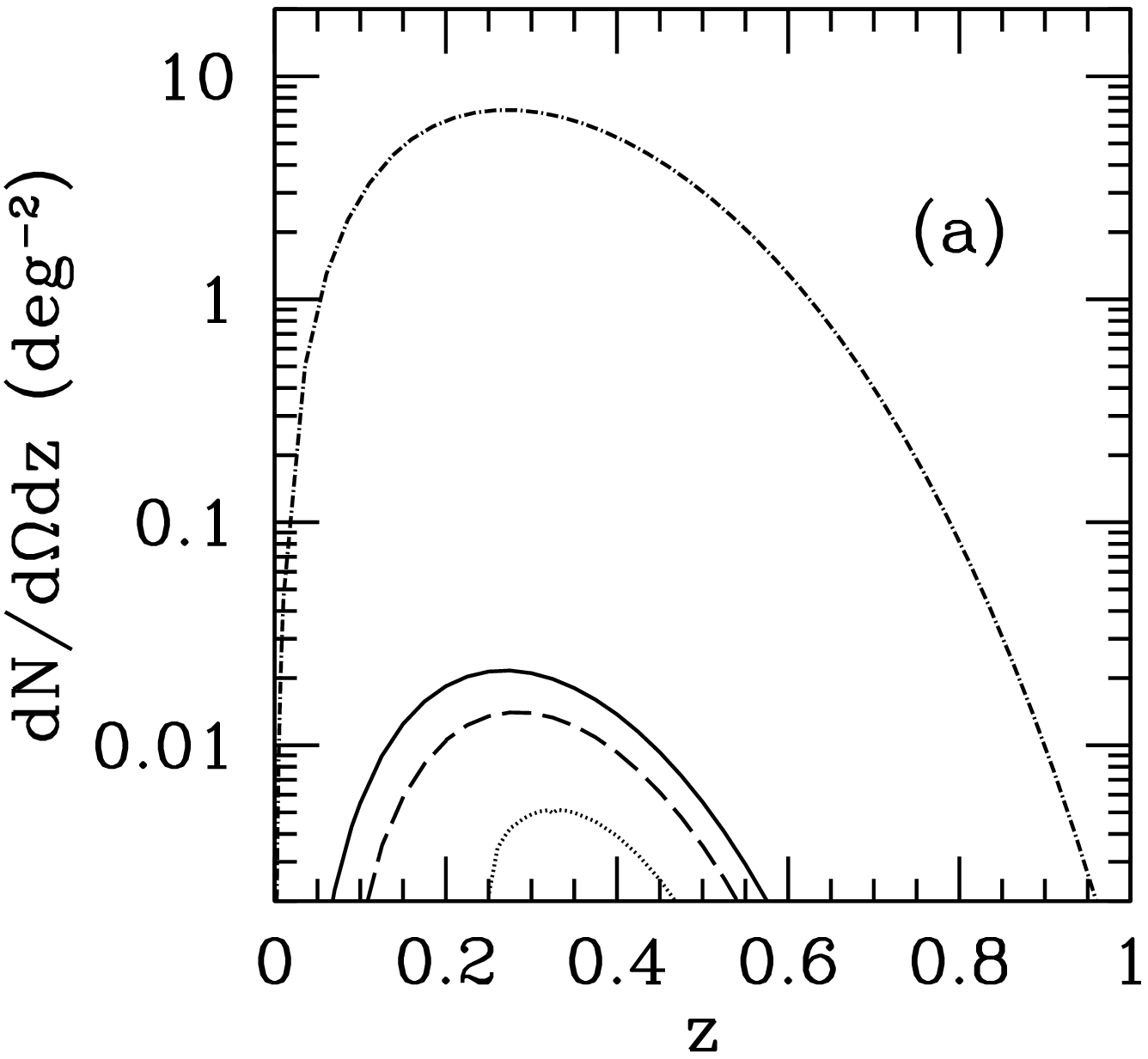}} \hspace{0.13cm}
\resizebox{5.7cm}{!}{\includegraphics{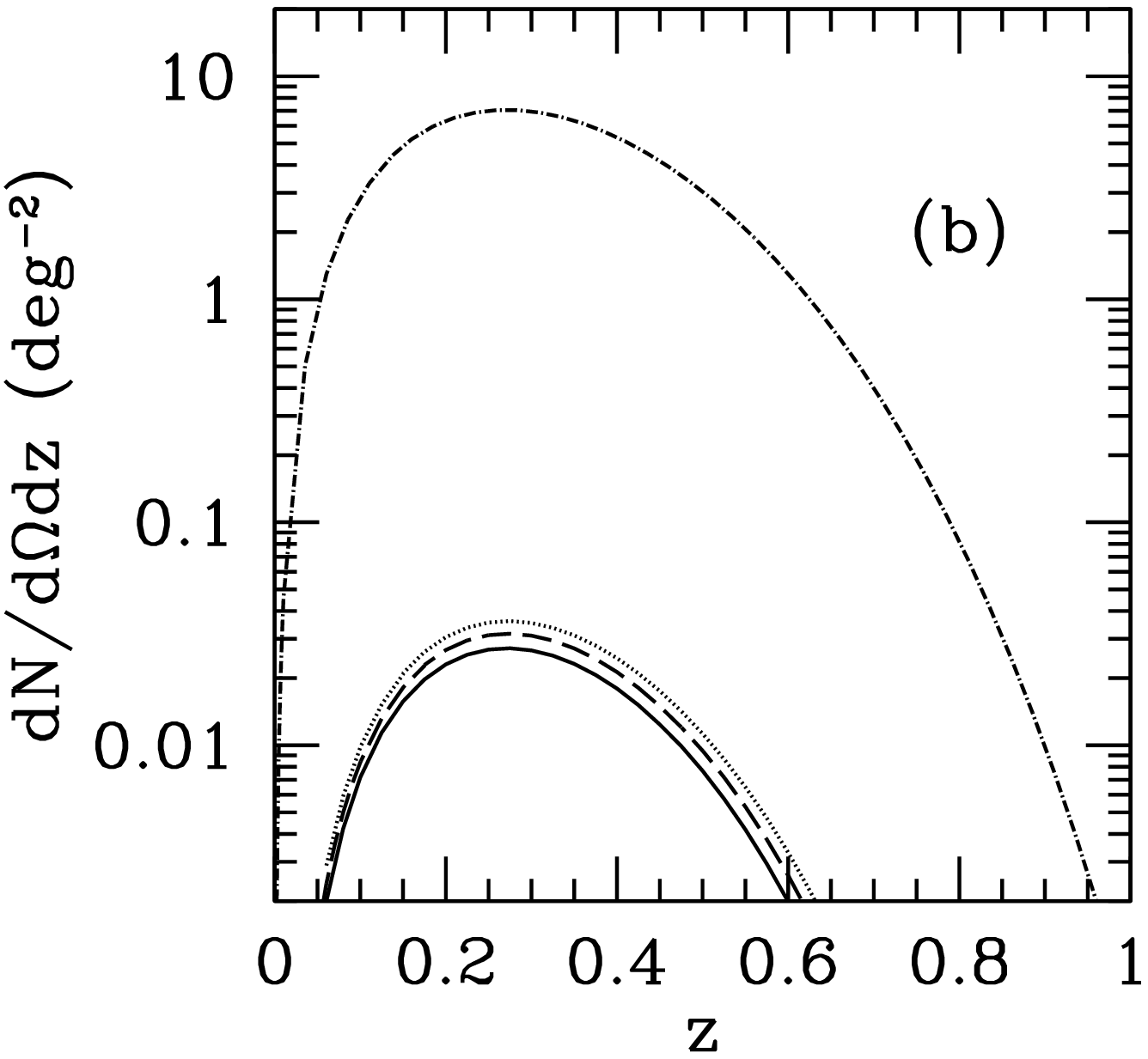}} \hspace{0.13cm}
\resizebox{5.7cm}{!}{\includegraphics{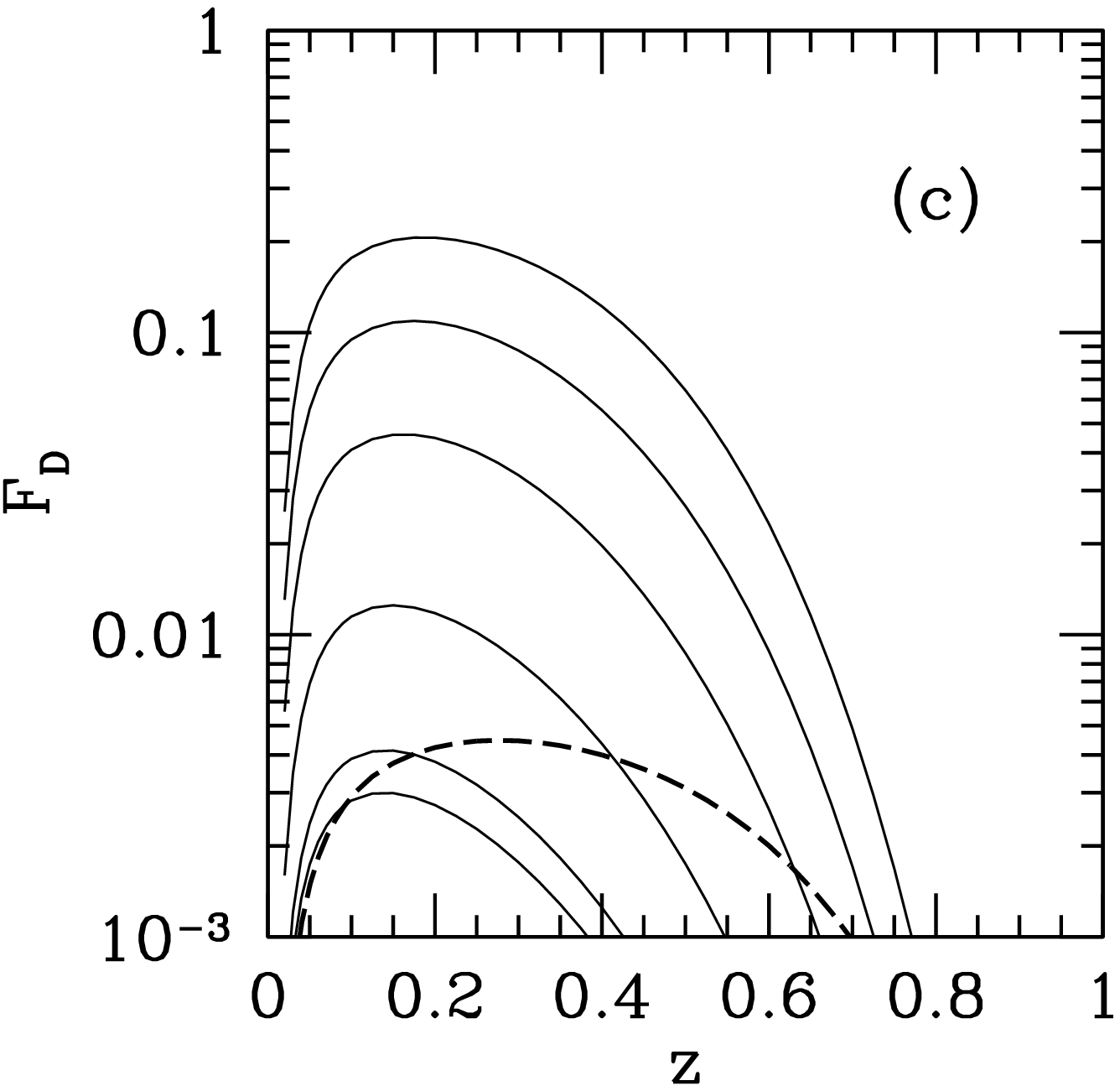}}
\end{center}
\caption{Panel $(a)$:  the differential number count of dark LPs for a hypothetical weak-lensing survey with background galaxy density $n = 30~\mathrm{arcmin^{-2}}$ and intrinsic ellipticity dispersion $\sigma_{\epsilon} = 0.3$.  We assume a fixed $f_{\mathrm{det}} = 0.1$.  The solid, dashed, and dotted curves correspond to $\alpha_{\mathrm{min}} = 0$, $0.2$, and $0.4$.  The dot-dashed curve shows the estimated differential number count of clusters.  Panels $(b)$ shows the same but with fixed $\alpha_{\mathrm{min}} = 0$ and $f_{\mathrm{det}} = 0.3$ (solid), $0.5$ (dashed), and $0.7$ (dotted).  Panel $(c)$:  the fraction of lenses that are dark LPs, $F_D$, as a function of redshift.  The dashed curve shows $F_D$ using our redshift-dependent cluster mass threshold $M_{\mathrm{cl}}(z)$.  From bottom to top, the solid lines correspond to constant $M_{\mathrm{cl}}$ of $8 \times 10^{13}$, $1$, $2$, $4$, $6$, and $8 \times 10^{14}~\Msun$.}
 \label{SurveyPL}
 \end{figure}

Similarly, the differential cluster number count is
\begin{equation}
\frac{\dd N_{\mathrm{cl}}(z)}{\dd z \dd \Omega} = \frac{\dd V}{\dd z \dd \Omega} \int_{M_{\mathrm{cl}}}^{\infty}{n_{\mathrm{h}}(M,z)~\dd M},
\label{clustercountsEQ}
\end{equation}
where, $n_{\mathrm{h}}$ is the halo mass function, and $M_{\mathrm{cl}}$ is the redshift-dependent mass detection threshold of the survey.  We use the Sheth-Tormen mass function for $n_{\mathrm{h}}$.  To estimate $M_{\mathrm{cl}}$ for a hypothetical survey using the filter (\ref{ApertureEQA}) with $n = 30 ~\mathrm{arcmin^{-2}}$ and $\sigma_{\epsilon} = 0.3$, we use a NFW profile truncated at the virial radius.  At a given redshift, we solve equation (\ref{StoNEQ}) for the mass that yields $S/N = 4$.  Since a wide range of $\tout$ and $x_c$ would be used in practice, we set $R_{\mathrm{out}}$ equal to the virial radius and $x_c$ equal to the cluster concentration in order to estimate the lowest detectable mass.  Using this method, we obtain detection thresholds of $M_{\mathrm{cl}} = 8.2 \times 10^{13}$, $1.1 \times 10^{14}$, $1.4 \times 10^{14}$, and $1.9 \times 10^{14}~\Msun$ at $z = 0.1$, $0.2$, $0.3$ and $0.4$ respectively.

The solid, dashed, and dotted lines in Figure \ref{SurveyPL} $(a)$ show the differential dark LP number count for $\alpha_{\mathrm{min}} = 0$, $0.2$, and $0.4$ respectively.  We assume a fixed $f_{\mathrm{det}} = 0.1$.  This means that all LPs under consideration have sub-haloes with less than a $10$ per cent chance of being detected via soft-band X-ray emission, assuming a flux limit of $10^{-14}~\mathrm{ergs~s^{-1}~cm^{-2}}$.  The plot shows that most dark LP detections correspond to objects between $z \sim 0.1 - 0.6$.  The dot-dashed line shows the differential cluster number count obtained using the $M_{\mathrm{cl}}(z)$ described above.

Note that the most efficient lenses are located at $z \sim 0.5$ (roughly halfway between the observer and the peak of the source distribution).  However, geometry is not the only factor affecting the differential number counts.  For both dark LPs and clusters, the growth of structure leads to an increase in the comoving number density at lower redshifts.  On the other hand, the differential number counts are suppressed in this regime due to the comoving-volume element.  At high redshifts, the lack of LSS and higher detection limits are responsible for suppressing the number counts in both cases.  The dark LP number counts decrease as $\alpha_{\mathrm{min}}$ increases because the low and high mass LPs with the flattest power-laws are cut out (see Figure \ref{dlensAPL}).

The solid, dashed, and dotted curves in panel $(b)$ correspond to $f_{\mathrm{det}} = 0.3$, $0.5$ and $0.7$ respectively.  We assume a fixed $\alpha_{\mathrm{min}} = 0$.  Raising $f_{\mathrm{det}}$ increases the dark LP number counts since overdensities with larger sub-haloes are included.  However, panel $(b)$ shows that this is a mild effect; the results are relatively insensitive to the choice of $f_{\mathrm{det}}$.

Panel $(c)$ shows the fraction of lenses that correspond to dark LPs,  

\begin{equation}
 F_D(z) = \frac{\int_{M_{\mathrm{min}}}^{M_{\mathrm{max}}}{n_D(M,z)~\dd M}}{\int_{M_{\mathrm{min}}}^{M_{\mathrm{max}}}{n_D(M,z)~\dd M} + \int_{M_{\mathrm{cl}}}{n_{\mathrm{h}}(M,z)~\dd M }}
 \label{ContaminationEQ}
\end{equation}
for fixed $\alpha_{\mathrm{min}} = 0$ and $f_{\mathrm{det}} = 0.5$.  From bottom to top, the solid curves correspond to constant $M_{\mathrm{cl}} = 8 \times 10^{13}$, $1$, $2$, $4$, $6$, and $8 \times 10^{14}~\Msun$.  The dashed curve was obtained using the redshift-dependent $M_{\mathrm{cl}}(z)$ described above.  Note that in the denominator of (\ref{ContaminationEQ}) we have neglected the shear selected counts due to other potential causes of dark lenses (ie LSS and allignment of intrinsic ellipticities).  Including these detections would decrease $F_D$.  We will discuss spurious signals in section \ref{DISCUSSION}.  Panel $(c)$ shows that even in our worst case estimate, where sub-haloes in LPs have up to a $50$ per cent chance of displaying observable X-ray emission and flat infall profiles are allowed, dark LPs only make up $\la 0.4$ per cent of lenses at at any given redshift (see the dashed curve in Figure \ref{SurveyPL} $(c)$).  

By definition, the redshifts of dark LPs are unobservable.  Hence, weak-lensing surveys will only be sensitive to their cumulative number.  We calculate the all sky number of dark LPs and clusters by integrating equations (\ref{darkLPnumbercountsEQ}) and (\ref{clustercountsEQ}) for $\alpha_{\mathrm{min}} = 0$ and $f_{\mathrm{det}} = 0.5$.  We obtain total dark LP and cluster numbers of $410$ and $108,303$ respectively.  We estimate that the former make up $\la 0.4$ per cent of the total number of lenses.  It is therefore unlikely that they will be responsible for a significant fraction of dark lens detections in future surveys.  Although we restrict our analysis to the aperture mass filter of \citet{Schirmer2004,Schirmer2007}, this result holds for other filters as well.  A filter that adds more weight to its outer regions would decrease the over-density required for a LP to create a weak lensing signal. However, the main factor suppressing dark LP abundances - the probability of finding such an over-density without large sub-halos - remains small. We emphasize that the dark LP abundance is generally small because it is highly unlikely to find an over-density large enough to create an observable lensing signal with sub-haloes small enough to escape X-ray detection.  

In accordance with current shear-selected samples,  we perform calculations using $n= 24~\mathrm{arcmin^{-2}}$ and $\sigma_{\epsilon} = 0.48$.  These parameters are more representative of the deepest exposures in surveys to date \citep[see][for example]{Schirmer2007}.  In this case we find that dark LPs cannot be detected by their weak-lensing signal.  Put in another way, there are no protoclusters with small enough virial masses that meet the $S/N = 4$ threshold.  In order to create a detectable weak-lensing signal, the virial masses have to be larger.  However, this means that the probability of X-ray detection increases.  Hence our model indicates that current shear-selected samples are not likely to contain a significant number of LPs that will escape deep ($\ga10^{-14}~\mathrm{ergs~s^{-1}~cm^{-2}}$) soft-band X-Ray searches.

\section{Discussion} \label{DISCUSSION}
 
We have developed an analytic model to determine whether LPs can account for a significant fraction of dark lenses.  In our model, a protocluster consists of a small virialised central region surrounded by infalling matter.  A dark LP corresponds to a cluster-scale mass concentration with group-sized ($\sim 10^{13}~\Msun$) virial mass.  The small virial mass results in a low probability of X-ray detection, while the large total mass yields a high aperture mass $S/N$.  As initially suggested by WK2002, these objects can potentially share the same observational properties as dark lenses.
  
For the LP mass distribution we used an idealized model consisting of an NFW profile inside of the virial radius and power-law, $\rho \sim r^{-\alpha}$, extending from the virial radius to the truncation radius.  In this case, the total $S/N$ is the sum of contributions from inside and outside of the virial radius.  In the case of small virial mass, the $S/N$ is dominated by the infall region.

Dark objects in shear-selected samples would likely be followed up with deep X-Ray searches.  In order to quantify the likelihood for a LP to display detectable X-ray emission, we used the analysis of \citet{Nord2008}.  We found that LPs with a low probability of being detected via their X-Ray luminosities (or equivalently low virial masses) must have large total masses ($M \sim 10^{15} \Msun$) and $\alpha$-values $\la 1$ to meet the aperture mass detection threshold.  Such infall regions may exist given the recent findings of \citet{Tavio2008}, who showed that the density profiles of haloes deviate from the NFW form beyond the virial radii, and display considerable scatter.  The abundance of these objects in N-body simulations has yet to be investigated.  Objects with the above characteristics would display rising $\left< \gamma_t \right>$ at larger radii.  A comparison of our results with the shear profiles of detected dark clumps is difficult due to the fact that their redshifts are unmeasurable by definition.  Hence, it is impossible to determine whether features in the shear profile occur at the appropriate radii.    

We have used the excursion set formalism to calculate the abundance of dark LPs.  In our approach, the number density of mass concentrations that are sufficiently overdense to meet the $S/N$ threshold is multiplied by the fraction that are unlikely to be detected in the soft X-ray band.  This subset of objects contains zero virialised sub-haloes with a probability $\ge f_{\mathrm{det}}$ of being detected via X-Ray emission.  In most cases of interest this fraction is extremely small, resulting in a suppression of dark LP abundances.  These results appear to be consistent with the average profiles derived in \citet{Tavio2008}, which indicate that infall regions typically do not contain enough mass to create dark lenses.

In section \ref{IMPLICATIONS}, we compared the differential number counts of dark LPs to ordinary clusters in a hypothetical shear-selected survey with source density $n = 30~\mathrm{arcmin^{-2}}$ and intrinsic ellipticity dispersion $\sigma_{\epsilon} = 0.3$.  In both cases, we found that most detections originate from objects at $z_l \sim 0.1 - 0.6$.  The dark LP number counts are generally $2-3$ orders of magnitude smaller than the cluster number counts.  

We varied the minimum allowed power-law index, $\alpha_{\mathrm{min}}$, to simulate scenarios in which flat infall power-laws are dynamically unlikely.  We found that dark LP abundances are highly sensitive to $\alpha_{\mathrm{min}}$, dropping rapidly with increasing $\alpha_{\mathrm{min}}$.  If infall regions typically fall off steeper than $r^{-1}$, then lensing contributions from the outskirts of dark LPs may be insufficient to meet the detection threshold.  We also varied $f_{\mathrm{det}}$ to explore the remote possibility that objects with large virialised sub-haloes could be detected as dark lenses.  We found that the differential number counts are relatively insensitive to $f_{\mathrm{det}}$, varying only by $\sim 20-30 $ per cent between $f_{\mathrm{det}} = 0.3$ and $0.7$.

Finally, we have calculated the fraction of lenses that correspond to dark LPs.  We found that they constitute $\la 0.4$ per cent of lenses at any given redshift.  Moreover, dark LPs account for $\la 0.4$ per cent of the total number of lenses in our hypothetical shear-selected survey.  We therefore concluded that dark LPs are too rare to be considered a plausible dark lens candidate.  

Our approach adds to initial work by WK2002 in several important ways.  The first is our use of the density profile (\ref{LPprofileEQ}), which provides a physical model for lensing protoclusters that takes into account deviations from the NFW form beyond the virial radius.  Such deviations have been recently pointed out in high resolution N-body simulations by \citet{Cuesta2008} and \citet{Tavio2008}.  An additional advantage of (\ref{LPprofileEQ}) is that we are able to quantify the $S/N$ contributions from virialised and unvirialised matter.    In contrast, the model of WK2002 does not allow one to quantify the lensing contribution from the central regions that meet the virialization over-density threshold.  Hence, in their approach it is possible to consider cases where the $S/N$ is dominated by the virial mass of the LP.  These cases typically occur when the total over-density of the LP is close to the virialization threshold.  Note that this difference is one of the reasons that our approach generally yields higher LP masses compared to the WK2002 results.  By forcing the virial region to be smaller in order to simultaneously minimize its lensing contribution and avoid X-ray detection, larger total masses are required to meet the $S/N$ threshold.  

We also point out that the fraction of mass in lensing overdensities does not correspond to the fraction contained in dark objects.  Many of these overdensities contain large virialised sub-haloes.  In practice, these cases would correspond to true cluster detections since follow up X-ray searches would be sensitive to these sub-haloes.  By incorporating the halo statistics of \citet{CasasMiranda2002}, our approach only counts overdensities without large virialised sub-haloes.  This key difference accounts for the lower dark LP abundances that we obtained compared to WK2002.  

Following the work of \citet{ReblinskyBartelmann1999,White2002,Hamana2004,Hennawi2005,Pace2007,Fan2007}, it is more likely that dark detections will correspond to false peaks resulting from: 1) LSS along the line-of-sight.  In this case, the $S/N$ is due to projected mass; it cannot be associated with a single isolated structure.  2) the random or correlated alignment of intrinsic galaxy ellipticities.  These alignments alone can lead to spurious detections, especially in shallow surveys.  However, it is also possible they can boost peaks that correspond to smaller mass concentrations \citep{vonderLinden2006,Fan2007}.  Owing to an artificially high $S/N$, these detections can be misinterpreted as dark lenses.
 
Using ray-tracing through stacked snapshots of cosmological N-Body simulations, \citet{Pace2007} tested the performance of the \citet{Schirmer2004} filter used above (referred to as OAPT in their paper).  By removing individual lens planes of haloes that may be associated with a particular $S/N$ peak, \citet{Pace2007} were able to separate true detections from spurious ones.  A true detection is associated with a cataloged cluster in the N-body simulation; a spurious detection remains when lens planes of individual candidates are removed.  They found that the OAPT filter yields spurious detection fractions $\la 20~(25)$ per cent at $S/N = 4$ for source redshifts of $z_s = 1~(2)$.  For larger aperture sizes, this fraction decreases only mildly at higher $S/N$ thresholds.  In addition, they point out that these spurious detections are indistinguishable from true detections in a $S/N$ map.  Therefore, it is likely that LSS accounts for at least some of the dark lenses reported in the literature.

Note that a dark LP would likely be counted as a ``true" detection with the algorithm of \citet{Pace2007}.  The removal of the lens plane containing the dark LP would significantly diminish the signal observed in the $S/N$ map.  In addition, since the CVR would be cataloged in the N-body simulation, the detection might be associated with this small-mass halo.  Therefore, it would be instructive to determine what causes the lensing enhancement of small-mass detections in studies such as \citet{Pace2007}.

Finally, we point out that the intrinsic galaxy ellipticities were randomly oriented in \citet{Pace2007}.  The effect of \emph{correlated} alignment of intrinsic ellipticites on the number of false detections was not taken into account.  As \citet{Fan2007} points out, galaxy formation is sensitive to the local environment.  One would therefore expect the  orientation of a galaxy to at least be correlated with its closest neighbors.  Such alignments can increase the number of false detections in convergence $\kappa$-maps significantly.  \citet{Fan2007} showed that including this source of noise can increase the likelihood of false detection due to intrinsic ellipticities in a given field.  This increase can affect whether intrinsic ellipticities can be ruled out in a dark lens detection.  Future numerical studies on false peaks in weak lensing surveys should investigate this important possibility.  

While the analytic model presented in this paper provides important insight into why dark LPs should be extremely rare, it is limited by several key issues.  The first is the simplistic density profile (\ref{LPprofileEQ}), which neglects the effects of anisotropy and substructure on the $S/N$.  A more detailed analysis should incorporate these properties, which are expected to have a significant effect on the lensing signal.  Secondly, since the excursion set formalism does not yield any information about the density profiles of individual trajectories, it is impossible to rigorously determine whether objects meet the $S/N$ threshold.  The best we can do in our analytic approach is to assume that objects above the derived over-density threshold can be detected.  In addition, we have used a simple model for the X-ray luminosities of protoclusters in order to estimate the probability of detection.  In reality this is a highly complicated problem with many caveats that can only be addressed numerically.  Lastly, our model does not include galaxy overdensities.  We assume that objects with low virial masses also escape optical selection.  Future studies should focus on whether LPs display low galaxy overdensities as well.  Each of the above issues would be ideally addressed in a high-resolution N-body simulation containing a baryonic component.  Our model provides a starting point for more detailed investigations on the characteristics of simulated protoclusters and their impact on shear-selected samples.  

\section*{Acknowledgments}

The authors thank Nevin Weinberg for helpful discussions.  We also thank the anonymous referee for helpful suggestions.

\bibliographystyle{mn2e} 
\bibliography{LensingProtoclusters}

\appendix{}

\section{The first crossing distribution of the weak lensing barrier} \label{APPENDIXA}

In this section, we obtain the probability of piercing the absorbing barrier, 

\begin{equation}
\delta_l = \left\{ \begin{array}{ll}
	 B_a S + A_a & \mbox{$ S_1 \le S \le S_2$} \\
 B_b S + (B_a - B_b)S_2 + A_a & \mbox{$S_2 <S \le S_3$}, \end{array} \right.
\label{barrierEQ}
\end{equation}
at a scale $S$.  The quantity $\delta_l$ represents an approximation to the minimum linear over-density required to create an aperture mass signal-to-noise ratio of 4.  Here, $S_1 = \sigma^2(M_{\mathrm{max}})$ and $S_3 = \sigma^2(M_{\mathrm{min}})$ are the mass scales corresponding to the maximum and minimum detectable LP masses discussed in the section \ref{LPsDL}.  $S_2$ is the mass scale at which $\delta_l$ transitions from slope $B_a$ to $B_b$.      

Fortunately, the problem can be greatly simplified by the fact that a trajectory beyond a given scale $S$ is independent of the path leading up to $S$. We do not want to count trajectories that simultaneously cross the virialization threshold, $\delta_v$, at $S < S_1$ and $\delta_l$ at $S_1 \le S \le S_3$.  These correspond to lensing overdensities within larger collapsed objects.  To avoid this problem, we use 

\begin{equation}
Q_{\mathrm{PS}}(S_1,\delta_1) = \frac{1}{\sqrt{2 \pi S_1}}\left\{ \exp\left(-\frac{\delta_1^2}{2 S_1}\right) 
	- \exp\left( -\frac{(2 \delta_v - \delta_1)^2}{2 S_1}\right) \right\}
\end{equation}
as the probability density for starting at the origin and ending at $\left\{S_1,\delta_1 \right\}$.  Hence, for example, the probability density for a trajectory starting at the origin with intermediate and end points of $\left\{S_1,\delta_1 \right\}$ and $\left\{S,\delta \right\}$ respectively, where $S_1 < S \le S_2$, is $Q_{\mathrm{PS}}(S_1,\delta_1) Q^{a}(S,\delta|S_1,\delta_1)$.  Here, $Q^{a}(S,\delta|S_1,\delta_1)$ is the conditional probability density for trajectories starting at $\left\{ S_1,\delta_1\right\}$, where $S_1 < S \le S_2$.  The superscript denotes that the barrier parameters in the regime $S_1 < S \le S_2$ are to be used.  The general form for this probability density is given by equation (\ref{ConditionalQ}) in appendix \ref{APPENDIXB}.  Similarly, if $S_2 < S \le S_3$ then the probability density is $Q_{\mathrm{PS}}(S_1,\delta_1) Q^{a}(S_2,\delta_2|S_1,\delta_1) Q^{b}(S,\delta|S_2,\delta_2)$, with the additional intermediate point $\left\{S_2,\delta_2 \right\}$.  Summing over intermediate points yields 

\begin{equation}
Q(S,\delta) = \left\{ \begin{array}{ll}  \int_{-\infty}^{B_a S_1 + A_a}{d\delta_1~Q_{\mathrm{PS}}(S_1,\delta_1) Q^{a}(S,\delta|S_1,\delta_1)} & \mbox{$ S_1 < S \le S_2$} \\ & \\ 
	 \int_{-\infty}^{B_a S_2 + A_a} \int_{-\infty}^{B_a S_1 + A_a}d\delta_2~d\delta_1~Q_{\mathrm{PS}}(S_1,\delta_1) Q^{a}(S_2,\delta_2|S_1,\delta_1) Q^{b}(S,\delta|S_2,\delta_2) & \mbox{$ S_2 < S < S_3$} \end{array} \right.	
\label{Qeq}
\end{equation}
The first crossing distribution $f_S(S,\delta_l)$ of  (\ref{barrierEQ}) for trajectories starting at the origin is

\begin{eqnarray}
\label{fcEQ}
f_S(S,\delta_l) & = & -\frac{1}{2}\left[ \frac{\partial Q}{\partial \delta}\right]_{-\infty}^{\delta_l(S)} \label{dminFCeq} \\  \nonumber & = &  
\left\{ \begin{array}{ll}  \int_{-\infty}^{B_a S_1 + A_a}{d\delta_1~Q_{PS}(S_1,\delta_1) f^{a}_S(S,\delta_l|S_1,\delta_1)} & \mbox{$ S_1 < S \le S_2$} \\ & \\ 
	 \int_{-\infty}^{B_a S_2 + A_a} \int_{-\infty}^{B_a S_1 + A_a}d\delta_2~d\delta_1~Q_{PS}(S_1,\delta_1) Q^{a}(S_2,\delta_2|S_1,\delta_1) f^{b}_S(S,\delta_l|S_2,\delta_2) & \mbox{$ S_2 < S < S_3$} \end{array} \right.	
\end{eqnarray}
where $f^a_S(S,\delta_l|S_1,\delta_1)$ and $f^b_S(S,\delta_l|S_2,\delta_2)$ are obtained from equation (\ref{ConditionalFCeq}).  

\begin{figure}
\begin{center}
\resizebox{8.0cm}{!}{\includegraphics{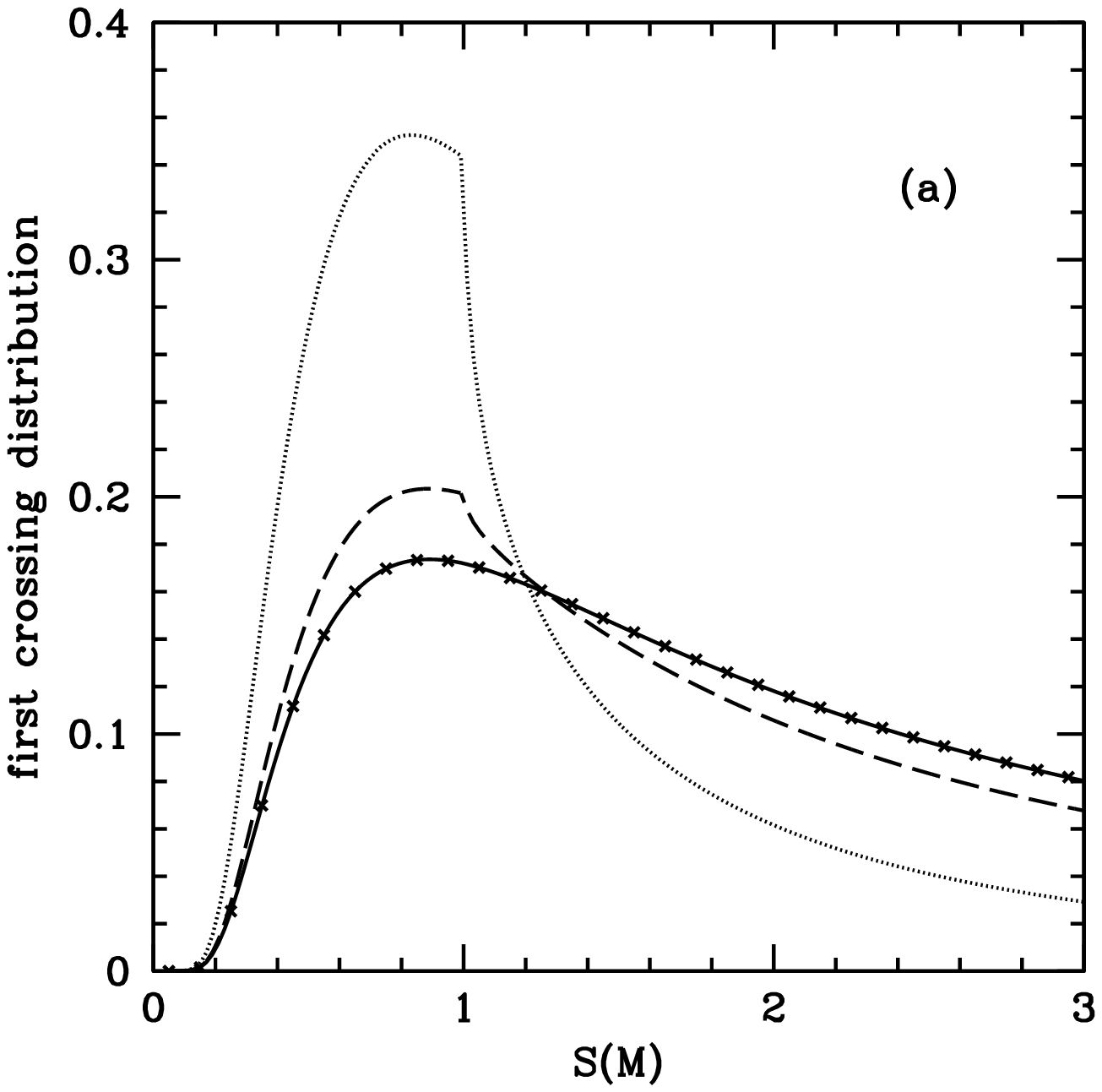}} \hspace{0.13cm}
\resizebox{8.0cm}{!}{\includegraphics{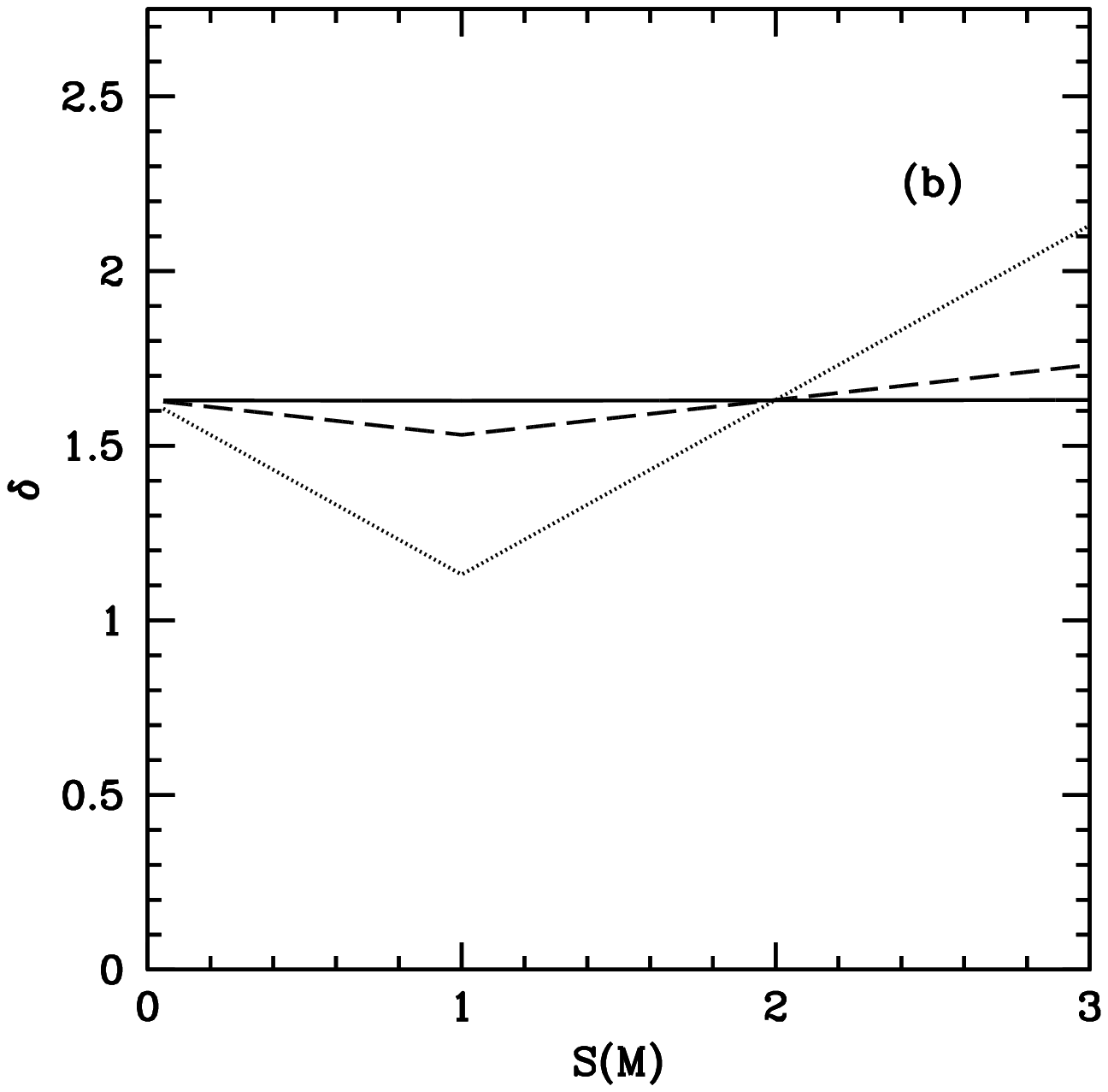}} 
\end{center}
\caption{Panel $(a)$:  comparison of equation (\ref{fcEQ}) to the PS first crossing distribution (shown with crosses) at $z = 0$.  The corresponding absorbing barriers are shown in panel $(b)$.  The crosses are overlaid on the solid curve in panel ($a$) because equation (\ref{fcEQ}) produces the same results as the PS case when the barrier (\ref{barrierEQ}) is approximately equal to the scale-independent virialisation threshold, $\delta_v$.  For the barriers that dip downward, it is more likely for trajectories to be absorbed leftward of the corner.  This depletion results in a smaller first-crossing probability rightward of the corner.}
 \label{FCexamplePL}
 \end{figure}

To illustrate the characteristics of (\ref{fcEQ}), we compare it to the PS first crossing distribution at $z = 0$ (depicted with crosses) in Figure \ref{FCexamplePL} $(a)$.  For the latter we use $\delta_v = 1.63$, which is obtained by applying the spherical collapse model to the virialisation threshold $\delta^{\mathrm{NL}} = 200\rho_c/\rhoBR - 1$.  The solid, dashed, and dotted curves correspond to (\ref{fcEQ}) with barrier parameters $B_a = -B_b = 0.001$, $-0.1$, and $-0.5$.  For all curves, we assume $S_1 = 0.1$, $S_2 = 1.0$, $S_3 = 5.0$ and $A_a =\delta_v$.  Panel $(b)$ shows the corresponding barriers.  

The kinks in panel $(a)$ correspond to the mass scale $S_2$ at which the two linear barriers meet in (\ref{barrierEQ}).  The solid curve shows that, in the limit where (\ref{barrierEQ}) is approximately constant with a value of $\approx \delta_v$, equation (\ref{fcEQ}) is equivalent to the PS form.  As the absorbing barrier dips down, it is more likely for trajectories to be absorbed in the $S_1 < S \le S_2$ regime.  In this case, the number of available trajectories to pierce the $S > S_2$ side of the barrier is depleted.  This effect can be observed in panel $(a)$ as a decrease in the first-crossing probability for $S > S_2$.

\section{Conditional first crossing distribution for the linear absorbing barrier}
\label{APPENDIXB}

In this section we obtain the first-crossing distribution of the linear absorbing barrier $\delta= B S + A$ for trajectories starting at $\{S_0,\delta_0\}$.  Solutions to the problem in which trajectories start at the origin can be found in \citet{Sheth1998} and \citet{McQuinn2005}.

For completeness, we summarize the diffusion equation approach taken by \citet{McQuinn2005} to obtain the general solution for $Q(S,\delta)$.  The probability density obeys the diffusion equation

\begin{equation}
\frac{\partial Q}{ \partial S} = \frac{1}{2} \frac{\partial^2 Q}{\partial \delta^2}
\end{equation}
with boundary condition $Q = 0$ for $\delta = B S + A$.  We utilize the linear transformation $y = B (\delta - B S)$ and $x = S - S_0$ to obtain

\begin{equation}
\frac{\partial Q}{ \partial x} = \frac{B^2}{2} \frac{\partial^2 Q}{\partial y^2} + B^2 \frac{\partial Q}{\partial y} 
\label{TransDiffEq}
\end{equation}  
with $Q(x,y= B A) = 0$.  Assuming a solution of the form $Q(x,y) = f(y) g(x)$, the problem is reduced to solving two ordinary differential equations, $g' = \lambda g$ and $(B^2 / 2) f'' + B^2 f' = \lambda f$.  The general solution can be written as an integral over the parameter $\lambda$,

\begin{eqnarray}
Q(x,y) & = & \int_{-\infty}^{-B^2/2}{d\lambda~h(\lambda) \exp^{-y + \lambda x} \left( c_1(\lambda) \exp^{i \tau y} + c_2(\lambda) \exp^{-i \tau y} \right)} \label{gensol} \\ \nonumber
& & + \int_{-B^2/2}^{\infty}{d\lambda~h(\lambda) \exp^{-y + \lambda x} \left( c_1(\lambda) \exp^{w y} + c_2(\lambda) \exp^{-w y} \right)}, 
\end{eqnarray}
where $\tau = \sqrt{2 |\lambda|/B^2 - 1}$ and $w = \sqrt{1 + 2 \lambda/B^2}$.  The condition $Q(x,y = B A) = 0$ cannot be satisfied simultaneously by both terms in equation (\ref{gensol}).  Moreover, the second term does not converge upon applying the above condition. Discarding the second term and rewriting the solution in terms of $\tau$ yields \citep{McQuinn2005}
 
 \begin{equation}
Q(x,y) =  \int_{0}^{\infty}{d\tau~h(\tau) \exp^{-y - B^2~(\tau^2+1) x/2} \sin\left\{ \tau (y - B A)\right\}}.
\label{LinearQ}
\end{equation}
We now apply the initial condition $Q(0,y) = |B|~\delta_D(y-y_0)$, where $\delta_D$ is the Dirac delta function, to equation (\ref{LinearQ}) to obtain $h(\tau) = 2 |B|~e^{y_0}~\sin\left[\tau (y_0-BA)\right]/\pi$.  Integrating yields the conditional probability density

\begin{equation}
Q(x,y|0,y_0) =  \frac{1}{\sqrt{2 \pi x}}\exp\left[ -\frac{B^2 x}{2} - y + y_0\right]\left\{ \exp\left[ \frac{-(y - y_0)^2}{2 B^2 x} \right] - \exp\left[ \frac{-(y+y_0-2A B)^2}{2 B^2 x} \right] \right\}.
\label{ConditionalQ}
\end{equation} 
The first crossing distribution can be obtained from equation (\ref{ConditionalQ}) using

 \begin{eqnarray}
f_x(x,A B|0,y_0) & = 
\left\{ \begin{array}{ll}  -\frac{d}{dS}\int_{\infty}^{B A}{Q(x,y|0,y_0) \frac{dy}{B}} = -\frac{B}{2} \left[\frac{\partial Q}{\partial y}\right]^{B A}_{\infty}& \mbox{$B < 0$} \\ & \\ 
	 -\frac{d}{dS}\int_{-\infty}^{B A}{Q(x,y|0,y_0) \frac{dy}{B}} = -\frac{B}{2} \left[\frac{\partial Q}{\partial y}\right]^{B A}_{-\infty} & \mbox{$ B > 0$,} \end{array} \right.	
\end{eqnarray}   
where the second set of equalities follow from using equation (\ref{TransDiffEq}). For $B<0$ and $B>0$, we obtain
 
\begin{equation}
f_x(x,A B|0,y_0) =  \frac{(A B - y_0)}{B \sqrt{2 \pi x^3}}\exp\left[ -\frac{(A + B x - y_0/B)^2}{2x}\right].
\label{ConditionalFCeq}
\end{equation} 

\end{document}